\documentclass[12pt]{article}
\usepackage{amssymb}
\usepackage{graphicx}
\usepackage{epsfig}
\usepackage{wrapfig}
\usepackage{amsfonts}
\usepackage{axodraw}

\newcommand{\lapproxeq}{\lower .7ex\hbox{$\;\stackrel{\textstyle
<}{\sim}\;$}}
\newcommand{\gapproxeq}{\lower .7ex\hbox{$\;\stackrel{\textstyle
>}{\sim}\;$}}
\newcommand{\stackdown}[2]{\lower 1.4ex\hbox{$\;\stackrel{\textstyle{#1}}
{\scriptstyle{#2}}\;$}}
\newcommand{\be}{\begin{equation}}
\newcommand{\ee}{\end{equation}}
\newcommand{\beq}{\begin{equation}}
\newcommand{\eeq}{\end{equation}}
\newcommand{\bea}{\begin{eqnarray}}
\newcommand{\eea}{\end{eqnarray}}

\newcommand{\D}{\displaystyle}
\newcommand{\elle}{\ell\hspace{-0.16cm}/}

\newcommand{\pslush}{p\hspace{-0.16cm}/}

\newcommand{\nd}[1]{/\hspace{-0.6em} #1}

\makeatletter

\def\slash{\@ifnextchar[{\fmsl@sh}{\fmsl@sh[0mu]}}
\def\fmsl@sh[#1]#2{%
  \mathchoice
    {\@fmsl@sh\displaystyle{#1}{#2}}%
    {\@fmsl@sh\textstyle{#1}{#2}}%
    {\@fmsl@sh\scriptstyle{#1}{#2}}%
    {\@fmsl@sh\scriptscriptstyle{#1}{#2}}}
\def\@fmsl@sh#1#2#3{\m@th\ooalign{$\hfil#1\mkern#2/\hfil$\crcr$#1#3$}}
\makeatother

\def\beq{\begin{equation}}
\def\eeq{\end{equation}}

\def\I#1{{\rm Im}\,#1}

\catcode`\@=11 

\def\lsim{\mathrel{\mathpalette\@versim<}}
\def\gsim{\mathrel{\mathpalette\@versim>}}
\def\@versim#1#2{\vcenter{\offinterlineskip
    \ialign{$\m@th#1\hfil##\hfil$\crcr#2\crcr\sim\crcr } }}

\def\t1{{\tilde 1}}

\def\slash#1{#1\hskip-6pt/\hskip6pt}

\def\to{\rightarrow}

\begin{document}

\begin{titlepage}

\begin{flushright}
FTUV-031118
\end{flushright}

\begin{centering}

{\Large   Novel Phases and Old Puzzles in QED$_3$ and related models}
\vspace{0.4cm}

{\bf N.E.~Mavromatos$^{a,b}$} and  {\bf J. Papavassiliou$^{b}$}

\vspace{0.4cm}

$^a$ {\it King's College London, University of London, Department of Physics,
Strand WC2R 2LS, London, U.K.}

$^b$ {\it Departamento de F\'isica T\'eorica and IFIC, Universidad de Valencia,
E-46100, Burjassot, Valencia, Spain.}

\vspace{0.4cm}

{\bf Abstract}

\end{centering}

In  this  review we  discuss  novel  phases  (quantum critical  points
occurring only for $T=0$) in (2+1)-dimensional Abelian gauge theories,
which  may serve  as  prototypes  for studying  the  physics of  nodal
excitations in (underdoped)  high temperature superconductors.  We pay
particular  attention  to describing  the  existence  of novel  phases
related  to the  anomalous breaking  of certain  symmetries, including
unconventional  superconducting  properties.   Although  some  of  the
results are  rather old,  we present them  here from a  physical
perspective not discussed  previously  in the  literature.  In  this
respect we also explore the (in)applicability of some non perturbative
approaches on the  existence of a critical number  of fermion flavours
for chiral symmetry breaking in our specific models, which is an issue
closely related to the existence of the novel phases mentioned above.

\end{titlepage}

\section{Introduction}

Relativistic  gauge   field  theories  in   (2+1)-dimensions  play  an
important  r\^ole  in attempts  to  understand  the  dynamics and  the
associated  phase  diagrams  of  planar  doped  
antiferromagnets~\cite{sodano},  and
through this to shed light  on the still elusive microscopic theory of
high  temperature superconductivity~\cite{Dorey:1990sz,kovner,
slaveboson,recentslaveboson,herbut}.  The relativistic  nature  of the
pertinent   excitations   arises   when  considering   the   effective
(continuum) field theory that governs  the degrees of freedom near the
nodes of  the Fermi surface  or points where the  superconducting gaps
vanish.  Indeed, high  temperature superconductors  are experimentally
known to be strongly type  II d-wave superconductors, which have their
superconducting gaps vanishing at  some points in momentum space.  One
popular  approach to  the  physics of  high  $T_c$ is  to linearise  a
spin-charge  separating  theory~\cite{anderson} 
of  holon-spinons  around such  nodes,
resulting  in  relativistic  gauge  theories of  fermions  and  bosons
coupled to gauge fields.  This  type of theories provide a description
of  the  effective  interactions  in  the  ground  state  between  the
fundamental degrees of freedom; in this picture the physical electrons
are  not  considered  as  fundamental particles,  being  instead  {\it
composites} (bound-states) of spinons and holons~\cite{anderson}.

There are  two main approaches  to this problem, distinguished  by the
spin and  statistics properties of these  electron constituents: ({\bf
i}) the \emph{slave-fermion} approach~\cite{slavefermion,Dorey:1990sz},
in which the holons are viewed
as Dirac  fermions, electrically charged,  and the spinons  as bosonic
neutral  fields,  and  ({\bf  ii})  the  \emph{slave-boson}  
approach~\cite{slaveboson},
according to which the spinons  are viewed as neutral fermions and the
holons  as charged  bosons.  The  two approaches  are  supposed to  be
physically   equivalent,   by   means   of   appropriate   non-Abelian
bosonisation in a path integral continuous formalism~\cite{marchetti}.  
However, within
each approach  there are various models considered  in the literature,
which  may not  be  physically equivalent.  The associated  continuous
effective theories depend crucially on  the way the continuum limit is
taken.   This is  an  important  point, often  ignored  in the  recent
literature  on the  subject.  The upshot  of  the present  work is  to
critically  examine such  models  from  the point  of  view of  recent
non-perturbative   results  on   the  symmetry   structure   of  three
dimensional gauge theories, and argue on the existence of \emph{novel}
quantum  critical phases; the  latter, although  known from  the early
literature on three-dimensional gauge  theories, seem not to have been
taken  into account in  the recent  (condensed-matter) studies  on the
subject.

In this  work we  shall not give  a detailed description  of condensed
matter  results,  but rather  concentrate  on  the  comparison of  the
various  effective continuum gauge  models which  have been  argued to
describe the dynamics of  nodal excitations in doped antiferromagnetic
materials.  Even though  the  details of  the  microscopic theory  are
important for arriving at specific models they will not be crucial for
our  analysis.   We  refer  the  interested  reader  to  the  relevant
literature for  details on these  issues.  For our purposes  we shall
simply  associate  the  dynamics  of  the  spin-charge  separation  in
condensed matter systems to  effective continuum field theories, which
are  variants of three  dimensional quantum  electrodynamics (QED$_3$)
with 2N flavours of two-component Dirac fermions. To that end we shall
adopt the simplest version of the spin-charge separation leading to an
Abelian statistical gauge interaction between spinons and holons.

In what follows  we will discuss the various  models, with emphasis on
those  physical  properties which  indicate  the  appearance of  novel
phases, not discussed  in the recent literature. In  addition, we will
critically examine  the possibility  of restricting or  excluding such
phases by resorting  to recent non-perturbative arguments~\cite{appel}
related to the chiral  symmetry structure of (three-dimensional) gauge
theories.  In particular, we  shall argue  that, due  to a  variety of
reasons,  the constraints of  \cite{appel} are  not applicable  to the
models considered in this work.

The outline of the article is  as follows: In section 2 we discuss the
effective  field theories  of relevance  for the  underdoped  phase of
planar antiferromagnets. We put emphasis on the unconventional phases,
occurring either  at $T=0$  or at  most up to  very low  (mK) critical
temperatures, which  may be superconducting  in the (Landau)  sense of
having massless poles in the electric current-current correlator. Such
phases are characteristic of non  compact theories, and are absent for
compact  ones, where no  massless poles  appear in  the aforementioned
correlator.    We   discuss   such   phases  for   all   variants   of
three-dimensional  Abelian  gauge theories  of  interest to  condensed
matter.  In section 3 we  discuss critically some
non-perturbative  conjectures of \cite{appel}  on the  chiral symmetry
breaking  structure of gauge  theories.  If  this type  of conjectures
(which  are shown  to  hold in  most  four-dimensional examples)  were
valid, this  would challenge the existence  of quantum-critical phases
for some of the models, but not for all (specifically the $\tau_3$-QED
model  of \cite{Dorey:1990sz} escapes  this constraint).   However, we
present a variety of reasons  as to why the inequality of \cite{appel}
may not be applicable to the  class of models that we consider.  These
reasons range from infrared  (IR) infinities, which invalidate some of
the  counting arguments,  to potential  misconceptions related  to the
ultraviolet   (UV)  behavior   of  QED$_3$   and  in   particular  its
(non)-display of asymptotic freedom.  This last notion, as well as the
issue of the existence of a  critical number of fermion flavours for a
breaking of  chiral symmetry  are examined from  the point of  view of
Schwinger-Dyson (SD)  equations in the context of  a specific approach
in section  4. In  particular, the appropriate  definition of  what we
call a  (dimensionless) effective  charge in QED$_3$  is given,  and a
non-linear SD equation is derived for the semi-amputated vertex, which
constitutes  the   natural  non-perturbative  generalization   of  the
aforementioned concept.  The  resulting integral (SD) equation governs
the  dynamical  evolution  of  the  coupling, and  can  be  solved  by
transforming  it  to  an   equivalent  differential  equation  of  the
Emden-Fowler type.   Its solution indicates  that unless some  type of
mass  is   dynamically  generated  the   coupling  displays  unbounded
oscillatory behaviour in the IR.   When the coupling obtained from the
semi-amputated vertex is  inserted into a standard gap  equation it is
found  that chiral  symmetry does  take place  and a  fermion  mass is
indeed generated dynamically.  A  subsequent comparison of our results
to  those  obtained using  large-$N$  techniques  is  made.  Our  main
conclusion from this SD analysis is that, although there exist regions
of  allowed  values  of  this  effective charge  for  chiral  symmetry
breaking to occur, nevertheless there is no restriction on the allowed
number of fermion flavours.   Conclusions and outlook are presented in
section  5.  Finally  in an  Appendix we  review briefly  some  of the
salient facts of 
the pinch technique (PT)~\cite{Cornwall:1982zr,Cornwall:1989gv}.
The latter is a systematic
method for constructing gauge-independent off-shell Green's functions,
and could in principle lead to a manifestly gauge-invariant truncation
of the SD series.  

\setcounter{equation}{0}
\section{Novel effects in nodal liquids}

In this section we present arguments pointing towards the existence 
of anomalous superconductivity and new quantum critical points
in nodal liquids at $T=0$.

\subsection{An effective field theory approach}

Relativistic     superconductors     in    (2+1)-dimensions,     where
superconducting  properties are due  to some  anomalous graphs  in the
effective  continuum  gauge  theory,  have been  first  considered  in
\cite{Dorey:1990sz},   as  a   way   of  explaining   parity-invariant
high-$T_c$ planar  superconductors with  either nodal points  in their
Fermi  surface,  or  Fermi   surfaces  consisting  of  small  isolated
spherical pockets  (or points). The  basic idea behind such  models is
spin-charge   separation~\cite{anderson},  according   to   which  the
fundamental  degrees   of  freedom  in  the  ground   state  of  doped
antiferromagnets (AF) are not the ordinary electrons, but two kinds of
confined  constituents,   one  with  possibly   fractional  spin,  but
electrically  neutral,  termed   \emph{spinon},  and  the  other  with
electric charge, and possibly spinless, termed \emph{holon}.

The interest  in nodal  liquids has been  recently revived in  view of
experimental   results  confirming   the  $d$-wave   nature   of  high
temperature  superconductors,  as well  as  the  fact  that the  Fermi
surface of these materials in  the so called underdoped phase consists
of four  nodal points. In the  recent theoretical literature  a lot of
works  have  appeared linearising  the  spin-charge separating  theory
about  such nodes, in  an attempt  to understand  the dynamics  of the
underdoped cuprates.

There  are in  general two  formal  approaches to  describe the  field
theory   of    spinon   and   holons:   ({\bf    i})   the   so-called
\emph{slave-boson}  approach~\cite{slaveboson} treats  the  spinons as
neutral fermions and the holons  as charged bosons, and ({\bf ii}) the
so-called                                          \emph{slave-fermion}
approach~\cite{slavefermion,Dorey:1990sz}, which treats the spinons as
neutral bosons and the holons  as charged fermions.  It turns out that
the field-theoretic  description of the dynamics around  a nodal point
in the  Fermi surface is  best accomplished through an  effective {\it
relativistic  gauge field  theory},  where the  gauge field  expresses
statistical   effective  interactions   between  spinons   and  holons
(essentially  spin-spin interactions,  due to  the magnetic  origin of
superconductivity       in       the      spin-charge       separation
approach~\cite{anderson}).

\begin{figure}[htb]
\centering
\epsfig{file=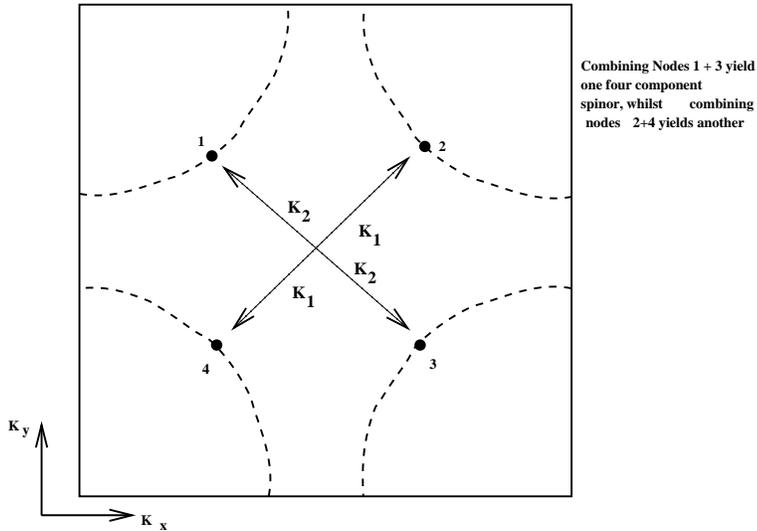, width=0.7\textwidth}
\caption{{\it In the QED$_3$ models of \cite{herbut,recentslaveboson}
the four component spinors (spinons) of the continuum
effective theory are constructed by combining nodes 
along the diagonal; their flavour index expresses the 
existence of two such pairs. In this scheme there are variations 
as to the precise nature of the components of the four-component
spinors, which may lead to important physical differences~\cite{herbut}.}}
\label{nodal1}
\end{figure}

The various models existing in the literature 
may be classified in two major categories, depending on the 
nature of their fundamental excitations: 

\begin{itemize} 

\item{(A)}       In       the       slave-boson      approach       of
\cite{recentslaveboson,herbut},  the  effective  nodal theory  of  the
spinon part  is nothing but QED$_3$, 
coupled to charged  boson degrees of freedom, corresponding
to the holons.  In this model the gauge interactions are assumed to be
Abelian,  non-compact   interactions.  The  way   the  four  component
continuous spinors,  representing the  spinons, are constructed  is by
means of  combining nodes as  shown in figure \ref{nodal1}.  There are
variations on  the precise  construction of the  effective microscopic
degrees of  freedom, which may lead to  important physical differences
of the resulting continuum theory, especially as far as the insulating
phase of the materials is concerned~\cite{herbut}.  We shall not dwell
upon these differences here, given that they have little or no bearing
on  the issue  of interest,  namely  the appearance  of novel  quantum
critical  points; the  latter  seem  to be  common  to all  approaches
mentioned above,  and will be the  focal point of  the discussion that
will follow.

\begin{figure}[htb]
\centering
\epsfig{file=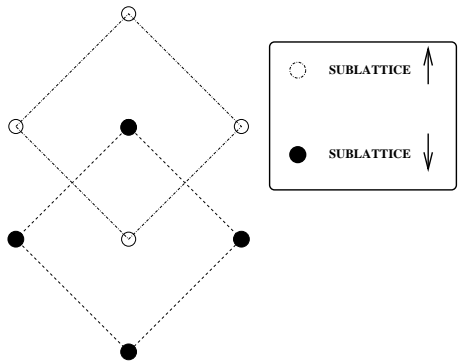, width=0.4\textwidth}
\caption{{\it  In the $\tau_3-QED$ model of \cite{Dorey:1990sz}, 
the four component 
spinors of the continuum theory (holons) 
are constructed by combining lattice points, and
then there is a ``colour'' $\tau_3$ index coming 
from the Antiferromagnetic sublattice structure.  }}
\label{nodal}
\end{figure}

\item{(B)}    In   the    slave-fermion    approach   considered    in
\cite{Dorey:1990sz}  the  four-component   spinors  of  the  resulting
effective  nodal  theory represent  the  electrically charged  holons,
while  the spinons  are $CP^1$  bosons (magnons).  The  four component
holons    are    constructed    by   combining    appropriately    the
antiferromagnetic  sublattice   structure,  as  indicated   in  figure
\ref{nodal}.   The   resulting  continuum  theory   is  the  so-called
$\tau_3$~QED$_3$  model, in  which the  statistical gauge  field 
couples
with  opposite   coupling  to  the   two  `colours'  of   the  holons,
representing   holon  excitations  in   each  sublattice.   The  gauge
interactions  of  the model  may  be  actually  embedded in  the  full
non-Abelian SU(2) gauge  group structure~\cite{farakos}, which results
in  the possibility  of having  
compact Abelian  gauge 
groups~\footnote{For an alternative SU(2) formulation of the 
doped $t-J$ model, within the slave-boson framework, see
\cite{plee}.}. As we
shall discuss below, there are important physical differences, between
compact and non-compact cases.

\end{itemize}

The  two   approaches  mentioned   above  {\it  are   not}  physically
equivalent,  due to  the different  ways by  which the  four component
spinors are  constructed, but also due  to the distinct  origin of the
statistical  gauge fields.  One  {\it should  not confuse}  this issue
with the one in which  the slave-boson and slave-fermion Ans\"atze are
related  by  means of  non-Abelian  bosonization  in  a path  integral
approach~\cite{marchetti}. Each one of the models in (A) and (B) above
can  be {\it  bosonised}  or {\it  fermionised}  accordingly, but  the
resulting models will {\it not}  connect (A) to (B).  They will simply
be the bosonised (or fermionised) versions  of (A) or (B). This can be
easily  understood  from the  fact  that,  in  the approach  (B),  for
instance, the spinons and holons couple to the statistical gauge field
with {\it opposite  couplings}, due to the AF  sublattice structure, a
feature which will persist  upon bosonization. This would imply simply
that  the [spinor  spinons]  of the  fermionised  approach derived  as
in~\cite{marchetti}    will    be     different    from    those    of
\cite{herbut,recentslaveboson},  whose coupling  with  the statistical
gauge field does not have the above-mentioned AF ``color'' structure.

Despite this  inequivalence, however, {\it both}  approaches are still
characterised  by anomalously-broken  symmetries,  leading to  quantum
critical points at $T=0$, although the broken symmetries are different
in  each  case.  The  purpose  of the  following  two  subsections  is
precisely to analyze these phases in some detail in the context of (A)
and (B).  Even  though the possible existence of  such phases has been
advocated  for  quite  some  time~\cite{kovner,Dorey:1990sz},  in  our
opinion  their  relevance  has   not  been  fully  recognized  by  the
condensed-matter community.

We shall  argue in this work  that their existence  may have important
implications for the entire phase  diagram of the high-$T_c$, and more
general the doped  AF.  We commence our discussion  from the case (B),
which historically  was the first  (2+1)-dimensional {\it relativistic
model} of  gauge field theories  to be used  in the context  of planar
doped  AF,  generalising  a  (1+1)-dimensional  spin  chain  model  of
\cite{shankar}.

\subsection{$\tau_3$-QED in the slave-fermion spin-charge separation }

We   start   from   a    brief   description   of   the   results   of
\cite{Dorey:1990sz},  which  utilizes  the slave-fermion  approach  to
spin-charge separation.   In this  case, the electron  operators are
written as~\footnote{For the most part of 
this work  we restrict ourselves  to Abelian
spin-charge separation  Ans\"atze for brevity; one can carry  over the
discussion to non abelian cases,  where the statistical gauge group is
non abelian~\cite{farakos}, we shall make some comments on the physics
of the non-abelian case later on in the article.}:
\begin{equation}
c_{i,\alpha} = \psi_i {\overline z}_\alpha~, \quad \alpha = 1,2 
\label{sf}
\end{equation} 
where $\psi_i$ ($i$ is a Lattice site index)
denotes the spinless holons, which are  
Grassmann number on the lattice
carrying electric charge but no spin, 
and $z_\alpha$ are $CP^1$ magnons, 
representing the spinons, with $\alpha = 1,2$ a spin SU(2) index. 
There is of course the constraint of having at most one electron per lattice site
\begin{equation} 
\psi^\dagger_i \psi_i + \sum_{\alpha =1}^{2}{\overline z}_i z_i =1~, 
\quad ({\rm no~sum~over~i}) 
\label{constraint} 
\end{equation}

We first give a brief but comprehensive 
description of the continuum construction presented in 
\cite{Dorey:1990sz}, where 
the interested may find more details. 
In that work, 
the antiferromagnetic nature of the underlying lattice defines two 
``colours'' of two-component fermions;  
one obtains two additional flavours
because of the usual lattice doubling (c.f. figure \ref{nodal}). 
In terms of nodes these are the degrees of 
freedom around each node in the construction of \cite{Dorey:1990sz}.
In contrast, in the constructions of \cite{herbut,recentslaveboson},
which can also be considered in the context of $\tau_3$-QED~\cite{ams} 
one combines the two nodal points 
along the diagonal of the appropriate Fermi graph, to construct one species
of four component spinor, and then one duplicates the species 
by taking into account the combination of the 
other two nodes, as indicated in figure \ref{nodal1}. 
However, in that case one should couple the 
two so-obtained four-component spinors to the gauge fields
with opposite statistical couplings
in order to get the $\tau_3$-QED model,
which allows the statistical gauge interactions to be viewed as forming a 
subgroup of the gauged spin SU(2). 
This is the important 
physical difference compared to 
the construction of \cite{herbut,recentslaveboson},
where one does not distinguish the sublattices, as far as the statistical
gauge interactions are concerned. In that case, the latter 
are not directly related (embedded) to the gauged spin SU(2) group.

The effective continuum relativistic nodal (2+1)-dimensional 
lagrangian reads~\cite{Dorey:1990sz}:
\begin{eqnarray} 
{\cal L} = \gamma {\rm Tr}|(\partial _\mu - i a_\mu\tau_3)z|^2 + 
{\overline \Psi } (i\gamma^\mu \partial_\mu + \gamma_\mu a^\mu \tau_3 
+ \frac{e}{c}\gamma^\mu A_\mu )\Psi + \dots 
\label{tau3}
\end{eqnarray} 
where $z$ are spinons (bosons), and $\Psi^c, c=1,2$ are four-component 
fermions, with a colour index associated with the antiferromagnetic 
nature of the underlying microscopic lattice $t-J$ model~\cite{Dorey:1990sz}.
The gauge field $a_\mu$ is 
an abelian field expressing the 
effective spin-spin interactions among the fundamental excitations of the model, 
and is different from the real (external) electromagnetic field $A_\mu$, to  
which the electrically charged holons $\Psi^c$ couple. Notice that,
as a result of the antiferromagnetic nature, there is a 
$\tau_3 =\left(\begin{array}{rr} 1 &  0 \\ 0 & -1 \end{array}\right)$ 
coupling
for the  $a$-field, but  an ordinary QED  coupling for  the $A$-field.
Moreover,  the  effective  ``speed  of  light''  of  the  relativistic
continuum theories  is the Fermi velocity  of the node,  which here is
taken to be  one. It is for this reason  that the real electromagnetic
speed of light , $c$ has been kept explicit in the $A$-$\Psi$ coupling
in (\ref{tau3}).  In realistic models it is estimated that $c=10^4$ in
units of the fermi velocity of  the nodes. This should be kept in mind
throughout  this work.  The  $\dots$ in  (\ref{tau3}) express  contact
interactions among the fermions. From them, the most relevant ones are
four-holon  interactions, which  in (2+1)-dimensions  are known  to be
\emph{relevant} in a Renormalization-group sense~\cite{fourfermi}.  We
shall come back to their r\^ole  in connection with the claims made in
\cite{appel} in this context in the next section.

\begin{figure}[htb]
\centering
\epsfig{file=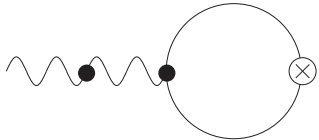, width=0.3\textwidth}
\caption{{\it The breaking of fermion number symmetry upon mass generation
    of fermions in the $\tau_3$QED$_3$ model of \cite{Dorey:1990sz}. The figure denotes
the matrix element of the fermion number current between the vacuum and one
`photon' state. The dark blobs denote loop corrections and the crossed-blob
denotes an insertion of the fermion number current.}} 
\label{anom}
\end{figure}

We now remark that the bare mass of the holons $\Psi^c$ 
is zero, but a \emph{parity conserving}  mass can be generated 
dynamically by means of the 
statistical gauge interactions $a_\mu$~\cite{app,Dorey:1990sz}. 
The parity conserving mass is energetically preferred  
due to a theorem in \cite{vafawitt} for vector-like theories. 
In the absence of external fields $A_\mu$, 
and in the \emph{massive phase of the spinons} 
$z$, which can thus be integrated out 
producing kinetic Maxwell terms  $\Pi (0) f_{\mu\nu}^2$ ($\Pi (k)$ is 
the one loop vacuum polarization), 
such a mass is responsible
for a spontaneous breaking (i.e. by the ground state of the system)
of the \emph{global} fermion-number symmetry, generated 
by the current $J_\mu = {\overline \Psi}^c \gamma_\mu \Psi_c $. The breaking is due to the \emph{anomalous} graph of fig. \ref{anom}, as discussed in detail 
in \cite{Dorey:1990sz}. The $\tau_3$ coupling is crucial to this effect.

The resulting matrix element is: 
\begin{equation} 
<a_\mu |J_\nu |0> \sim \frac{M}{|M|}\epsilon_{\mu\nu\rho} \frac{p^\rho}{\sqrt{p_0}}
\label{matrixelem}
\end{equation}
where $M$ is the mass of the fermion $\Psi$. 
The result (\ref{matrixelem}) is perturbatively \emph{exact}.

\begin{figure}[htb]
\centering
\epsfig{file=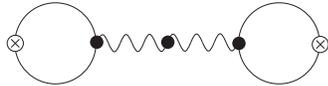, width=0.3\textwidth}
\caption{{\it In non compact gauge theories the anomalous graph of fig. \ref{anom} 
gives rise to  
massless poles in the current-current correlators,  
as a result of the exchange of the massless statistical 
gauge boson $a_\mu$; upon coupling the system to an  
external electromagnetic potential, 
this results in superconductivity 
(Landau criterion).}}
\label{anomsc}
\end{figure}

Upon coupling to external electromagnetic fields $A_\mu$, the 
anomalous matrix element (\ref{matrixelem}) (c.f. fig. \ref{anom}) 
leads to \emph{superconductivity} in the case of \emph{non compact}
gauge fields $a_\mu$, as a result of the existence of a massless
pole in the electric current-current correlator~\cite{Dorey:1990sz}
(c.f. fig. \ref{anomsc}):
\begin{equation} 
<0|J_\mu (k) J_\nu (-k)|0> = \frac{g^2}{\pi^2}
\left(\frac{k_\mu k_\nu}{k^2} - \delta _{\mu\nu}\right)
\left(1 + \Pi (k)\right)^{-1}
\label{matrixelemel}
\end{equation}
where $\Pi (k)$ is the one-loop vacuum polarization graph.

\begin{figure}[htb]
\centering
\epsfig{file=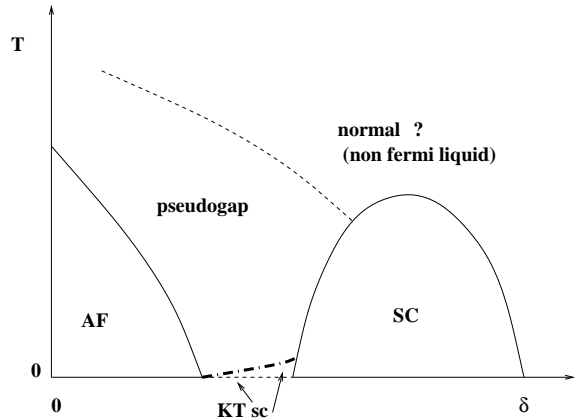, width=0.5\textwidth}
\caption{{\it The anomalous graph of fig. \ref{anom} 
results to a novel quantum critical phase, 
of Kosterlitz Thouless (KT) type (no local order parameter perturbatively)
strictly at $T=0$. Should the model of \cite{Dorey:1990sz} 
describe the underdoped cuprate phase, then, this 
may modify the phase diagram of high temperature 
superconductors, by connecting the optimal doping superconducting 
phase and the antiferromagnetic phases, as indicated 
by the dashed line. It must be stressed, though, 
that if the screening
of the magnetic field lines still occurs for finite temperatures
up to a critical $T_c^\star$ much lower than the optimal $T_c \sim 100$ K
of high $T_c$~\cite{Dorey:1990sz}, 
then the novel KT superconducting phase extends also
inside the pseudogap phase in a region bounded by such small $T_c^\star$, 
as indicated by the dashed dotted line.}}  
\label{modscphd}
\end{figure}

Strictly speaking 
this superconducting behaviour pertains only to zero temperatures 
$T=0$, since at \emph{finite} temperature the masslessness of the 
statistical gauge field $a_\mu$ disappears, as  a result of a 
plasmon mass in the longitudinal 
component $a_0$. 
In view of the above considerations therefore, the novel 
$T=0$ phase predicted here would result in a 
modification of the temperature-doping phase diagram 
of high-$T_c$, by a quantum critical line at $T=0$, as 
shown in figure \ref{modscphd}, 
should the $\tau_3$-model describe 
correctly the underdoped cuprate phase. 
It has been argued, though, in~\cite{Dorey:1990sz} 
that despite the plasmon longitudinal mass, 
the screening of the magnetic field (Meissner effect),
associated with the massless transverse 
component of the statistical gauge field $a_\mu$,  
is still valid at finite temperatures, up to a temperature
in which the nodal holon mass gap disappears.
Such a temperature has been estimated in \cite{farakosext}
to lie \emph{at most} on the $mK$ scale, i.e. much lower than 
the  optimal-doping critical temperature  of $100$ K 
of high-temperature superconductors.

\subsection{QED$_3$ in the slave-boson approach}

The issue we want to bring up in this subsection 
is that in the non compact gauge $a$-field case,
a corresponding $T=0$ anomalous behaviour characterises also the 
QED$_3$ models of \cite{herbut,recentslaveboson}, within the 
slave-boson approach, but the symmetry 
that breaks anomalously in this case is {\it not}
the fermion-number symmetry, but a version of chiral symmetry
(but {\it not} the standard chiral symmetry which breaks due to 
the parity-conserving mass generation for fermions). In this sense,
the quantum critical point does not imply superconductivity, but an 
anomalous behaviour within an AF phase. 
This quantum-critical phenomenon has been missed 
in the relevant recent 
literature~\cite{recentslaveboson,herbut}, because 
the analysis is restricted only to $T \ne 0$.  
It is our conjecture, therefore, that in QED$_3$ theories 
of the pseudogap phase of high-temperature superconductors 
utilizing non compact gauge fields, there is always an anomalous  
behaviour in the spin sector at $T=0$, 
which is realized via the loop graphs of 
figs. \ref{anom},\ref{anomsc}. By the way, 
we call this behaviour ``anomalous''
since at tree level the result is zero, and appears only at one loop.

Let us try to decipher this anomalous behaviour from studies of effective
QED$_3$ models of the slave-boson approach to spin-charge 
separation developed in~\cite{recentslaveboson,herbut}. According to the 
latter, the electron operators are written as:
\begin{equation}
c_{i,\alpha} = \xi_{i,\alpha} {\overline b}_i, \quad \alpha = 1,2 
\label{sb}
\end{equation} 
where $b$ are spinless bosons, 
carrying electric charge degrees of freedom,
and represent the holons, 
while $\xi_{i,\alpha}$ 
are electrically neutral fermions, carrying spin, which  
represent the spinons, with $\alpha = 1,2$ a spin SU(2) index.
There is of course the corresponding constraint (\ref{constraint})
again, expressed in terms of the new variables in (\ref{sb}). 

In \cite{recentslaveboson,herbut} an effective continuum 
field theory for $\xi$ has been constructed, which 
yields an interacting QED$_3$ model of two 
four-component spinors $\Psi$ constructed appropriately out of 
$\xi_{\alpha}$, $\alpha=1,2$, interacting with a \emph{non compact}  
statistical $U(1)$ gauge field, expressing spin 
frustration responsible
for nodes mixing. The lagrangian reads: 
\begin{eqnarray} 
{\cal L}_{QED_3} = -\frac{1}{4}f^2_{\mu\nu}(a)
+ \sum_{f=1}^{2}{\overline {\tilde \Psi}}_f (i\gamma^\mu \partial_\mu + \gamma_\mu a^\mu ){\tilde \Psi_f} + \dots 
\label{qed3}
\end{eqnarray} 
where ${\tilde \Psi_f}$ are electrically neutral 
four-component continuum spinors (spinons), and $f=1,2$ 
runs over node pairs as in figure \ref{nodal1}. 
The massive ${\tilde \Psi}$ phase is characterised first of all by 
a breaking of (global) \emph{chiral symmetries} 
generated 
by $\gamma_5 = i\left(\begin{array}{rr} 
0 & I \\I & 0\end{array}\right)$ and 
$\gamma_3 = i\left(\begin{array}{rr} 
0 & I \\-I & 0\end{array}\right)$, 
in a four-component notation for spinors with an even number of 
flavours, 
in a reducible representation
of the Dirac algebra in $(2+1)$-dimensions. 
Above,  $I$ denotes the $2\times 2$ unit matrix. These are global symmetries, 
whose breaking will result in 
the appearance of massless Goldstone bosons, 
to be discussed in the next section.

\begin{figure}[htb]
\centering
\epsfig{file=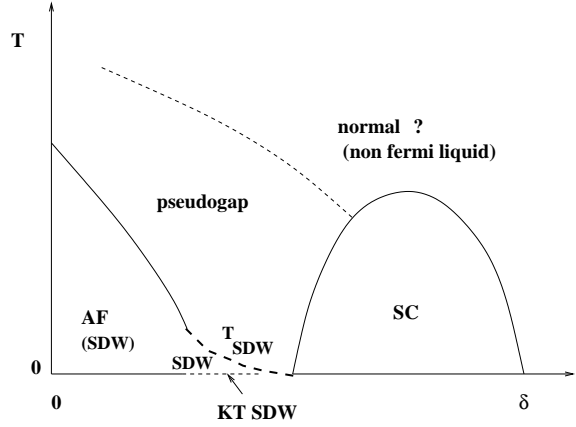, width=0.5\textwidth}
\caption{{\it In the $QED_3$ model of \cite{herbut,recentslaveboson}, 
the anomalous graph of fig. \ref{anom},
but for the chiral symmetry breaking current 
${\overline \Psi}{\tilde \tau}_3\gamma_\mu \Psi$, 
results to a novel quantum critical phase, 
of KT type (no local order parameter perturbatively)
which is a spin density wave (SDW) (weak AF) phase. 
This novel KT superconducting phase extends 
inside the pseudogap phase, 
and covers a region from the $T=0$ line
to a critical temperature which is much smaller 
than the critical temperature separating the pseudogap 
from the normal phase. For any $T \ne 0$ it becomes indistinguishable
from the conventional chiral symmetry broken phase of \cite{herbut},
but its $T=0$ anomalous behaviour, with a massless pole 
in the two-point correlator of the $\tau_3$-chiral symmetry currents, 
provides a distinctive discontinuous novel behaviour. Compare/contrast the 
situation with that in figure \ref{modscphd}, where one obtains 
an anomalous KT superconducting phase connecting 
the optimal doping superconducting (SC) phase with the AF phase. Here 
one obtains an anomalous 
SDW phase that connects the AF and (optimal doping) SC phases.}}  
\label{herbutfig}
\end{figure}

A dynamical mass generation (spin gap)
for the spinons ${\tilde \Psi_f}$  
has been argued in \cite{recentslaveboson,herbut} to be related with properties
of the pseudogap phase of high temperature superconductors. 
Most importantly, 
the massive spinon phase 
has been interpreted in \cite{herbut} as implying 
that the antiferromagnetic (AF) phase 
is immediately succeeded by the superconducting phase, as indicated
in figure \ref{herbutfig}, in the sense of a 
region of weak AF (spin density wave (SDW) phase) in which the critical
temperature for the new phase is much smaller than the pseudogap critical
temperature. This situation is opposite to the graph in figure \ref{modscphd}
of the model of \cite{Dorey:1990sz}, where the superconducting phase
(for non compact gauge fields) enters the pseudogap region until
the AF (Neel) state. The analysis of \cite{herbut} did a low temperature
analysis, attempting to generalize the conclusions down to the $T=0$ region. 
It is our purpose here to point out some important aspects 
of this region in QED$_3$, not discussed in \cite{herbut}, but existing
in the literature~\cite{kovner}, which like the $\tau_3$-QED case,
imply a discontinuity of the $T=0$ line (quantum critical line).

Indeed, apart from the conventional chiral symmetry in theories with even number of fermion flavours, discussed in \cite{herbut}, 
generated by the $4\times 4$ matrix $\gamma_3, \gamma_5$, 
there is \emph{another} 
symmetry of (\ref{qed3}) ~\cite{kovner,Dorey:1990sz}, 
whose breaking is realized in 
the so-called {\it Kosterlitz-Thouless mode}, 
without a (perturbative) local order parameter, and hence Goldstone bosons.
This symmetry is 
generated by the 
${\tilde \tau_3} = 
i\left(\begin{array}{rr} 
I & 0 \\0 & -I\end{array}\right)$
The corresponding current generator
is $J_\mu^{{\tilde \tau}_3} = \sum_{f=1}^{2}\overline{{\tilde \Psi }_f} 
{\tilde \tau_3 }\gamma_\mu {\tilde \Psi_f} $ 
and, as shown in \cite{kovner} its matrix element between the vacuum and the one photon state (as in fig. \ref{anom}) is non-zero in the massive 
$\Psi$ phase: 
\begin{equation} 
<a_\mu |J_\nu^{{\tilde \tau_3}} |0> \sim \frac{{\tilde M}}{|{\tilde M}|}
\epsilon_{\mu\nu\rho} \frac{p^\rho}{\sqrt{p_0}}
\label{tau3matrixelem}
\end{equation}
This also leads to $J_\nu^{{\tilde \tau_3}}$ 
current-current correlators with a massless pole for strictly a $T=0$ case,
according to the corresponding graphs of fig. \ref{anomsc} (c.f. (\ref{matrixelemel})).  
Our claim is that such a KT $T=0$  phase, with a massless pole in the 
respective correlator of the generating currents
defines a {\it novel} 
quantum critical 
point which corresponds to an unconventional AF phase in the model
of \cite{herbut}, or, in fact, any other QED$_3$ model 
(within the slave-boson approach) of doped AF.
As in the case of $\tau_3$-QED, the critical temperature at which the 
chiral-symmetry-breaking spinon mass disappears, is also a critical 
temperature for this anomalous behaviour. This temperature is much smaller
than the critical temperature separating the pseudogap from the 
normal phases in the phase diagram. The distinctive feature of this 
anomalous symmetry from the standard chiral symmetry, which is used
in the approach of \cite{herbut}, is precisely its $T=0$ behaviour,
where the current-current correlator  for the spinon current 
${\overline \Psi}{\tilde \tau}_3\gamma_{\mu} \Psi$ 
has a {\it massless pole} that disappears 
at any finite $T$, due to plasmon mass.  
This feature defines the $T=0$ as a quantum critical 
line, corresponding 
to a KT breaking of the $\tau_3$ chiral symmetry, defined above:
we may term this phase as KT SDW (c.f. figure \ref{herbutfig}).  
For any $T \ne 0$ the situation is like in \cite{herbut}, but the
$T=0$ quantum critical behaviour is distinct, and `discontinuous'
from the $T \ne 0$ case, 
and has been missed in the 
recent condensed matter literature~\cite{herbut,recentslaveboson}.

\subsection{Compact Gauge Theories} 

In the above analysis the statistical gauge field had been assumed {\it non compact}. However, the spin-charge separation Ansatz might 
be extended to a {\it non-Abelian} form~\cite{farakos}
by exploiting appropriately an (approximate) particle-hole symmetry of the 
$t-j$ Hamiltonian for underdoped cuprates: 
\begin{equation}
\chi _{\alpha\beta,i} = 
\psi _{\alpha\gamma,i}z_{\gamma\beta,i} \equiv \left(
\begin{array}{rr}
c_1 & c_2 \\
c_2^\dagger & -c_1^\dagger \end{array}
\right)_i = 
\left(\begin{array}{rr}
\psi_1 & \psi_2 \\
-\psi_2^\dagger & \psi_1^\dagger \end{array}
\right)_i~\left(\begin{array}{rr} z_1 & -{\overline z}_2 \\
z_2 & {\overline z}_1 \end{array} \right)_i 
\label{ansatz2}
\end{equation}
where the fields $z_{\alpha,i}$ obey canonical {\it bosonic} 
commutation relations, and are associated with the
{\it spin} degrees of freedom (magnons),  
whilst the fields 
$\psi _{a,i},~a=1,2$ have {\it fermionic}
statistics, and are assumed to {\it create} 
{\it holes} at the site $i$ with spin index $\alpha$ (holons).
The Ansatz (\ref{ansatz2}) 
has spin-electric-charge separation, since only the 
fields $\psi$ carry {\it electric} charge. It is a slave-fermion Ansatz
since the holons are fermionic.

The Ansatz 
has 
a {\it local} 
SU(2) symmetry, if one defines the transformation 
properties of the $z$ fields to be given by left multiplication
with the SU(2) matrices, and those of the $\psi ^\dagger_{\alpha\beta}$
matrices by the left multiplication
\begin{equation} 
     \xi_{\alpha\beta} \rightarrow h_\alpha^\gamma \xi_{\gamma\beta}
\label{localsu2}
\end{equation} 
In this representation, the gauge group 
$SU(2)$ is generated by the $2 \times 2$ Pauli matrices. 

The Ansatz (\ref{ansatz2}) possesses an 
{\it additional}  
local $U_S (1)$ `statistical' phase symmetry, which 
allows fractional statistics of the spin and charge 
excitations. This is an exclusive feature
of the three dimensional geometry. 
This is similar in spirit, although 
implemented in an admittedly less rigorous way, 
to the bosonization technique of the spin-charge 
separation Ansatz of ref. \cite{marchetti},
and allows the alternative possibility 
of representing the holes as slave bosons and  
the spin excitations as fermions. 

In addition, as a consequence of the fact that the fermions 
$\psi$ carry {\it electric charge}, one has an extra 
$U_{em} (1)$ symmetry for the problem. 

To recapitulate, 
the above analysis, based on the spin-charge 
separation Ansatz (\ref{ansatz2}) which allows 
spin flip,  
leads to the following 
local-phase (gauge) group structure for the doped large-$U$ Hubbard model:
\be
  G=SU(2)\otimes U_S(1) \otimes U_{em}(1) 
\label{group}
\ee
where the second $U_{em}(1)$ factor refers to electromagnetic 
symmetry due to the electric charge of the holes. 
This symmetry appears as a {\it hidden} symmetry of the
effective holon and spinon degrees 
of freedom obeying the Ansatz (\ref{ansatz2}). 

The statistics-changing $U_S(1)$ group is strongly coupled,
given that in the resulting effective lagrangian appears 
without a Maxwell Kinetic term, and hence it may be considered of 
(formally) {\it infinitely string} coupling $g_s \to \infty$.
In practice, the coupling is cutoff by an effective cutoff
which may be defined by the highest scale in the problem (e.g. the 
strong Hubbard coupling $U$ in Hubbard models with a strong repulsion 
$U \gg 1$ so as to implement the one-electron per site constraint, 
or the Heisenberg interaction $J$ in $t-J$ models~\cite{farakos}). 
Such strong Abelian groups may be responsible for dynamical mass
generation of fermions (holons). 

The effective continuum theory obtained from (\ref{ansatz2}) 
around the nodes of the underdoped cuprates consists of 
Dirac spinors, representing the holons, 
as well as $z$ magnons, interacting via the 
non-Abelian gauge interactions. The holons 
are coupled to the full Non-Abelian group G,
while the $z$ magnons, being electrically neutral, couple only 
to the $SU(2) \otimes U_S(1)$ subgroup. 

The dynamically generated mass for the holons is {\it parity conserving} 
as a result of the vector-like nature of the interactions in the simplest
model considered~\cite{vafawitt}. 
As discussed in detail in \cite{farakos}, such parity-conserving 
mass term {\it break} dynamically the SU(2) subgroup, due to the fact that
the parity conserving mass term for fermions transform like triplets
under this SU(2).   

As a result of the breaking $SU(2) \to U_c(1)$, 
two of the gauge bosons of SU(2) acquire heavy masses, and decouple
from the effective low-energy theory. 
One is then left with an effective
theory of massless degrees of freedom consisting of the unbroken 
subgroup associated with the $\tau_3=\left(\begin{array}{rr} 1 & 0 \\ 0 & -1\end{array}\right)$ generator of the SU(2) group. This model is then nothing else
but the $\tau_3$ QED model of \cite{Dorey:1990sz}, discussed above, with the 
important difference that now the statistical unbroken gauge group 
$U_c(1)$ is {\it compact}. 

Compact groups are known to have non-perturbative {\it monopole} 
configurations~\footnote{Recently, 
in the slave-boson framework, compact $QED_3$ theories
have also attracted attention from a different
perspective, specifically in connection with 
the role of monopoles on the confining phase, and their
implication on  
the Mott insulating phase of doped AF~\cite{herbut2}.}, 
which in (2+1)-dimensions are like {\it instantons}, having a point-like structure. Such monopole instantons are responsible in general 
for the generation of a small non-perturbative mass of the statistical 
photons of the $U_c(1)$ group~\cite{polyakov}:
\begin{equation} 
 {\rm photon~mass } \equiv m_\gamma \sim e^{-2S_0}  
\label{photonmass}
\end{equation}
where the above computations have been performed in a dilute instanton
gas, 
$S_0$ is the one-instanton action, and the factor of $2$ in the exponent
is due to the fact that the theory has fermions~\cite{ahw}. 
The presence of a non-perturbative photon mass, implies that in the compact
case there is {\it no longer} a massless pole in the electric current-current 
correlator (\ref{matrixelemel}). As a consequence, the 
KT $T=0$ superconducting quantum-critical line of the non-compact case 
is absent, and 
the AF phase is separated by the (optimal doping) superconducting one in the 
phase diagram of fig. \ref{modscphd} by a pseudogap phase, 
which in this case extends all the way down to $T=0$.

\setcounter{equation}{0}

\section{Counting degrees of freedom: a tricky business}

We would now like to 
address the issue of dynamical generation 
of the fermion mass {\it per se}
in all the above cases,
(A) and (B). As we have discussed, such a mass 
generation is crucial for the 
existence of the novel quantum-critical `anomalous' phases.
To this end, in this section we shall briefly review 
first the non perturbative arguments of~\cite{appel}, according to which 
dynamical mass generation can only occur in a theory 
with $N_f$ four-component spinors, with $N_f \le N_c = 3/2$,
showing that the previous large-$N$ treatment~\cite{app} 
overestimated $N_c$. 
If this result were true
in our case, this would mean that 
the above-described QED$_3$ 
model~\cite{herbut,recentslaveboson}, 
with two four-component spinors (spinons)
constructed as in fig. \ref{nodal1}, would never have 
a quantum phase with broken ${\tilde \tau_3}$ symmetry, since 
the spinons would be massless. 
However, as we shall discuss 
below, the arguments of \cite{appel} may not go through  
in (2+1)-dimensions for a variety of reasons.
At this stage we would like to mention, however, that 
even if such arguments were assumed to be valid in (2+1)-dimensional
gauge theories, nevertheless their application to 
the $\tau_3$-QED model~\cite{Dorey:1990sz}
{\it does not} select a critical number
for mass generation of the electrically charged fermions, leaving 
this task to a detailed SD analysis or other studies. 
Below we shall explain in detail why this is so, but we will also
provide arguments supporting the 
point of view that the analysis of \cite{appel} are not applicable to 
the (2+1)-dimensional QED$_3$ model either.

We commence our discussion 
by first going over
the new proposed constraint on strongly coupled field theories.
of \cite{appel} 
The constraint was based on 
a counting of massless degrees of
freedom between IR and UV fixed points, 
leading to inequalities between the respective 
values of the renormalized free energy density.
This  resembles, but is not identical to,  
the celebrated C-theorem
of Zamolodchikov for two-dimensional conformal field theories~\cite{zam}.
The conjectured constraint of \cite{appel}, 
has not been proven;
nevertheless, its validity has been verified in a number of physically
relevant examples. The conjecture of \cite{appel} implies a
reduction in the massless degrees of freedom as the theory flows under 
Renormalization Group (RG) from UV to IR fixed points.
In what follows we shall therefore examine the applicability of 
the conditions leading to the constraint for the above-described 
QED$_3$-like models of potential interest to the physics 
of doped antiferromagnets.

For instructive purposes, 
let us first state the form of the 
constraint of ref.~\cite{appel} and the main assumptions 
leading to it:
The main thrust of the constraint is associated with 
an inequality between the values of an extensive quantity,
such as the free energy, at the UV and IR
fixed points of \emph{asymptotically} free field theories. 
Specifically, it has been argued that the quantity  
$f_{\rm UV} = -{\rm lim}_{T \to \infty}\frac{{\cal F}}{T^4}\frac{90}{\pi
^2}$ where $
{\cal F}$ is the free energy of the system, 
and $T$ is the temperature, playing here the
r\^ole of a varying RG scale in the problem, 
counts the massless degrees of freedom in the UV, while 
a similar quantity: 
$f_{\rm IR} = -{\rm lim}_{T \to 0 }\frac{{\cal
F}}{T^4}\frac{90}{\pi ^2}$
counts the massless IR degrees of freedom of the system. 
The conjectured inequality  states that:
\begin{equation}\label{ineq}
f_{\rm IR} \le f_{\rm UV} \label{uvir}
\end{equation}
An important ingredient for the validity of Eq. (\ref{ineq})
is the
\emph{asymptotic freedom} of the theory in question. 
In fact, there are known examples
that do not satisfy the inequality  
if they exhibit a non-trivial UV fixed point~\cite{appel}.

The inequality (\ref{uvir}) has been applied to QED$_3$, 
which is
known to exhibit chiral symmetry breaking in the IR.
The theory has a dimensionfull coupling, and in \cite{appel}, it 
has been assumed
free in the UV.
According to \cite{appel}, this implies 
the following contributions of massless degrees of freedom to $f_{\rm UV}$
(in three space time dimensions): one
from the massless photon in (2+1)-dimensions, and $(3/4)~4N$ from $N$ 
four-component
free fermions. Hence one has at the UV: 

\begin{equation}
 f_{\rm UV} = 1 + 3N
\label{fuv}
\end{equation}

On the other hand, 
in the IR, there is a dynamical generation of
fermion masses, 
which implies a breaking of the global chiral symmetry of the massless
theory $U(2N) \to U(N)\times U(N)$.
In this case there are 2N$^2$ massless 
Goldstone bosons. The authors
of \cite{appel} deal with non compact QED$_3$, and hence the photon remains
massless in the IR, which implies the following contribution to $f_{\rm
IR}$ :

\begin{equation}
 f_{\rm IR} = 1 + 2N^2
\label{fir}
\end{equation}

{}From the inequality (\ref{uvir}), and (\ref{fuv}), (\ref{fir}), then, 
one obtains 
a \emph{critical number} of fermion
flavours for chiral symmetry breaking to take place; in particular 
$N < N_c = 3/2$, a result, which, if true, would imply  
that the early papers on dynamical mass generation~\cite{app} 
overestimated the
critical number, placing it in the region $3 < N_c < 4$.

Now let us go over these assumptions one by one:
\begin{itemize}
\item{(1)} \underline{Asymptotic freedom}:
In the context of perturbation theory  
QED$_3$ is super-renormalizable (i.e. it contains a finite 
number of divergent diagrams); this is a direct consequence of 
simple power-counting, given that the dimensionfull 
gauge coupling scales as $[M]^{1/2}$.
However, QED$_3$
is not asymptotically free, in the same way that 
the super-renormalizable  $(\phi^3)_4$ is not (this latter 
theory has a  dimensionfull coupling as well): 
 the coupling does \emph{not } go to zero for large momenta $p \to \infty$.
In fact, strictly speaking, the coupling does not display 
any energy dependence at all, because the corresponding $\beta$ function 
vanishes, i.e. $\beta_{QED_3} \equiv \mu \left(de/d\mu\right) = 0 $.
This is true simply because the vacuum polarization in $d=3$ is finite, and 
thus no wave-function renormalization for the photon is needed. 
The above renormalization group 
equation simply states that $e = e_0$, where $e_0$ is the 
initial tree-level coupling. Clearly such a theory cannot be 
asymptotically free (at least in a perturbative context), 
unless $e_0 =0$; but then this would imply that the 
theory is non-interactive (i.e. trivial) throughout. 
This situation is to be contrasted to what happens in the 
case of a {\it bona fide} asymptotically free theory, such as QCD$_4$.
In this case  $\beta_{QCD_4} = - b_0 g^3$ ($b_0 > 0)$; when solving 
this equation one obtains the usual logarithmic dependence of the 
coupling $g$ on $p$, which, in the limit $p \to \infty$, leads to 
an asymptotically non-interactive theory
(i.e. the value of the coupling goes to zero,
not just to some constant value).

Given the discussion above, in order to obtain some sort of 
non-trivial structure in the UV, one must study QED$_3$ outside 
the realm of standard perturbation theory.
In particular,  
arguments supporting asymptotic freedom of QED$_3$, 
and hence the counting in the UV
leading to (\ref{fuv}), are based on large $N$ treatments. 
Indeed, in such an
approximation~\cite{kondo} one defines a dimensionless effective running
number of flavours, which turns out to scale with the momenta as 
(in the non-local gauge):
\begin{equation}
{\rm (effective~coupling)^2/scale} = \frac{1}{N_{\rm run}(p)} \sim
\frac{1}{1 +\frac{g_0}{3}{\rm ln}p/\alpha}~;~ g_0 \equiv \frac{8}{\pi^2}
\label{asymfreed}
\end{equation}
An obvious drawback of the large $N$ treatment in this context is the 
fact that, 
even though the expansion is in principle systematic due to 
the presence of a parameter ($N$) which may be formally taken to 
infinity, in practice the desired effects are obtained 
by stretching the validity of this approximation 
into a range of $N$ values which one might consider as ``uncomfortably small''. 
In particular, an 
upper bound on the 
critical number of flavours $N < N_c \sim 3/2$ is obtained, beyond which 
chiral symmetry breaking is not possible. In this scheme 
the fundamental photon-fermion vertex is assumed to have the form  
$\Gamma_\mu = \gamma _\mu A^n(p)$, where
$A(p)$ is the wave function renormalization. For the case $N=2$ there is no
running and the model is not asymptotically free. For $N > 2$ the theory is
asymptotically free, while for $N < 2$ the theory 
behaves like QED$_4$ with a Landau pole. 
\footnote{In Kondo and Nakatani (first item in \cite{kondo}) 
the analysis was done
in the Landau gauge, while 
in the Kondo-Murakami paper (second item in \cite{kondo}) 
an improved analysis was performed
in the non-local gauge, where 
the vertex assumes the form
$\Gamma_\mu = \gamma_\mu G(p^2,q^2,k^2)$ but $A(p)$ can be set identically
equal to 1.} 

\begin{figure}[htb]
\centering
\epsfig{file=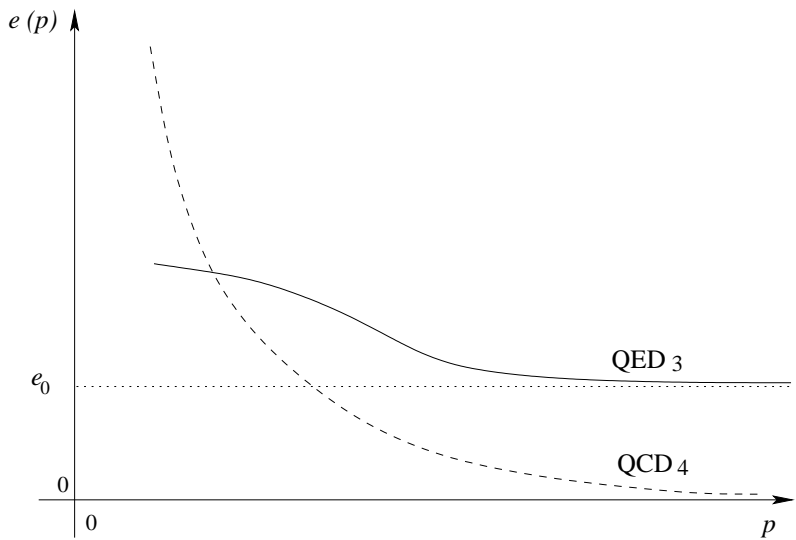, width=0.7\textwidth}
\caption{{\it The  dotted line represents
the constant (non-running) coupling obtained from the  $\beta_{QED_3}=0$ condition,
and the solid line is the running 
obtained through the effective charge of Eq.(\ref{effch3}). In both cases 
the theory is not asymptotically free, because the coupling saturates at a non-zero value.
For comparison, the $QCD_4$ coupling (perforated line) 
approaches asymptotically the value zero (asymptotic freedom)}}
\label{AF}
\end{figure}

{\it The Effective charge:} 

Given that, due to the super-renormalizability of  QED$_3$ one 
obtains no standard running for the coupling, one might be tempted to 
explore other field-theoretic alternatives. In particular one 
may ask what sort of ``running'' one obtains from the 
``effective charge'' in the case at hand. 
 
In QED$_4$ the infinite subset of radiative corrections summed in
the Dyson series generated by the
one--particle--irreducible vacuum polarization $\Pi_{R}(p^{2})$
defines an {\it effective charge}
which is gauge--, scale--, and scheme--independent
to all orders in perturbation theory:
\be\label{QEDalphaeff}
e^2_{\rm eff}(p^{2}) =
\frac{e_{R}^{2}}{1 + \Pi_{R}(p^{2})} =
\,\frac{e^{2}}{1 + \Pi(p^{2})}\,,
\ee
where we have used $e_{R}^{2} = (Z_{2}^{2}Z_{3}/Z_{1}^{2})e^{2}$ and $1 +
\Pi_{R} = Z_{3}(1 + \Pi)$
together with the QED Ward identity $Z_{1} = Z_{2}$ to write
$e^2_{\rm eff}(q^{2})$ purely in terms of bare quantities.
At $p^{2}/m_{f}^{2} \to\infty$, the effective charge
$e^2_{\rm eff}(p^{2})$ matches on to the {\em running coupling}
$\bar{e}^{2}(p^{2})$ defined from the renormalization group:
at the one--loop level,
\be
\lim_{p^{2}/m_{f}^{2}\to\infty} e^{2}_{\rm eff}(p^2) = \bar{e}^{2}(p^{2}) = 
\frac{e^2_{R}}
{1 - \frac{e^2_{R}}{12 \pi^2} n_{f}
\log(p^{2}/m_{f}^{2})}\,,
\ee
where $n_{f}$ is the number of fermion flavors.
What makes the effective charge a particularly useful
concept is its non-trivial dependence
on masses, through the analyticity properties of $\Pi (p^{2})$.
In particular,
when $p^2 > 4m_0^2$ the vacuum polarization
has imaginary part $\Im m \Pi(p^2)$ given by
\be
\Im m \Pi(p^2)  = 
\vartheta(p^2 - 4m_{0}^{2}) \frac{\alpha}{3}
\Bigg( 1 + \frac{2m_{0}^{2}}{p^2}\Bigg) 
\sqrt{1-\frac{4m_{0}^{2}}{p^2}} \,,
\label{IMP}
\ee 
By virtue of analyticity 
the real part $\Re e\Pi (p^2)$ 
may be reproduced from $\Im m \Pi(p^2)$ 
by means of a once--subtracted
dispersion relation 
Thus, for the one--loop contribution
of the fermion $f$, choosing the on--shell renormalization scheme,
\be\label{qeddisprel}
\Re e\Pi (p^2) =
\bigg(\frac{1}{\pi}\bigg) p^{2}\,
\int_{4m_{0}^{2}}^{\infty}ds\frac{\Im m \Pi(s)}{s(s-p^2)}
\ee
Finally, the one-loop $\Im m \Pi(p^2)$
is directly related, via the optical theorem, to the tree level
cross sections for the physical processes
$e^{+}e^{-}\rightarrow f^{+}f^{-}$, with  $f\neq e$, 
\be
\Im m \Pi(s) =  
\frac{s}{4\pi\alpha}
\,\sigma (e^{+}e^{-}\to f^{+}f^{-}) 
\label{OT}
\ee

The aforementioned properties of the 4-$d$ effective charge 
go through in the case of QED$_3$, with the appropriate 
adjustments due to the fact that no renormalization is needed
(for example, the 3-$d$ analogue of the dispersion relation in 
Eq.(\ref{qeddisprel}) needs no subtraction)\footnote{
The concept of the effective charge has been generalized 
in a non-Abelian context through the use of the PT 
\cite{Papavassiliou:1996zn,Papavassiliou:1996fn} .
Recently it has been proposed 
that the effective charges provide a natural framework for the 
reliable study of the impact of threshold effects 
of the unification of gauge couplings 
in Particle Physics models~\cite{Binger:2003by}}. 
In the case of QED$_3$ the   
dimensionless effective charge reads  
\begin{equation}
e^2_{\rm eff}(p) /\alpha = \frac{1}{1 + \Pi (p^2)} \label{pinched}
\label{effch3}
\end{equation}
where $\Pi(p^2)$ is the scalar co-factor of the 
one-loop vacuum polarization $\Pi_{\mu\nu}(p)= 
(p^2 g_{\mu\nu}- p_{\mu}p_{\nu})\Pi(p^2) $
in $d=3$; in particular, $\Pi(p^2) = - \alpha/8p$.  
The qualitative behavior of $e_{\rm eff}(p)$ is shown in 
fig.(\ref{AF}), solid line: For large $p$ the effective charge 
saturates to a {\it non-vanishing} value $e_0$, i.e. it does {\it not}
display asymptotic freedom. 

\item{(2)} \underline{IR infinities and the Mermin-Wagner theorem:}

The argument leading to the inequality (\ref{uvir}) is based on
a finite temperature extension. However, in two spatial dimensions at finite
temperature there are IR infinities as $T \to 0$, which invalidate the
well-defined nature of Goldstone bosons. Such IR infinities can be seen
in the effective IR cut-off dependence of the running flavour number at
any finite temperature $N = N({\rm ln}T)$ which stems from solving large-$N$
SD equations in the presence of an IR
cutoff~\cite{kondo,aitch,temp}. One may question the 
robustness of large-$N$ treatments,  but the
presence of IR infinities is probably indisputable.

The absence of Goldstone bosons, due to IR infinities (Mermin-Wagner
theorem), implies  the following counting of massless degrees of freedom in
the IR: $f_{\rm IR} = 1 $. Hence
in that case the inequality (\ref{uvir}) becomes formally (even 
assuming asymptotic
freedom):
\begin{equation}
f_{\rm IR} = 1 < f_{\rm UV} = 1  + 3N \quad \to N > 1/3 \label{irmerminwagner}
\end{equation}
hence we obtain no non-trivial 
information from this constraint. In fact the above
counting is formal, given that the IR infinities imply ln$T$ divergences
as $T \to 0$ in the effective action, and hence in $f_{\rm IR}$. 
One therefore has
to do the computation  explicitly to regularize these
divergences~\cite{temp,aitch}, which leads to IR-cutoff- 
(and hence temperature-) dependent critical number of fermion flavours $N_c(T)$.

\item{(3)} This and the following items are physical reasons why the
inequality (\ref{uvir}) does not apply to systems of interest in 
our case.

In most theories of condensed matter for high temperature superconductors there
are four fermion interactions present, 
in addition to the gauge minimal coupling
terms. It is well known that~\cite{fourfermi} four-Fermi theories in less 
than four dimensions are renormalizable (in a large $N$ treatment at least), 
and they
exhibit a non-trivial UV fixed point. This is sufficient to invalidate
the counting of degrees of freedom leading to $f_{\rm UV}$ (\ref{fuv}) in this
case, and hence the inequality (\ref{uvir}) for such systems.

\item{(4)} Another important physical system which may imply a different
critical number even if one accepts the inequality (\ref{uvir}) is the model
for high temperature superconductivity of \cite{Dorey:1990sz} known as
$\tau_3$QED$_3$, whose lagrangian is given in (\ref{tau3}). 
The important difference of this model from QED$_3$ (\ref{qed3})
lies in the original global symmetries. 
Namely, in the $\tau_3$QED model (\ref{tau3}), 
due to the opposite couplings of the two fermionic `colors' (each color is
a four component spinor), the symmetry that is broken is a global fermion
number symmetry $J_\mu = {\overline \psi}_{c,f}\gamma _\mu \psi_{c,f}$~\footnote{It must be stressed at this point that one may define a 
chiral-like symmetry in this case 
generated by $C\gamma_5$, where $C$ is the (statistical) charge conjugation 
operator, however due to the presence of the discrete operator $C$ 
this cannot be represented as a global $U(2N)$ chiral symmetry,
whose breaking leads to Goldstone bosons.
At any rate, this symmetry is broken explicitly in the specific model 
by the coupling 
of the charged fermions (holons) 
to the real electromagnetic potential~\cite{Dorey:1990sz}.}.
As discussed previously 
the symmetry is broken through the anomaly graph of figure \ref{anom}.
The figure denotes the S-matrix element
\begin{equation}
<a_\mu|J_\nu|0> ={\rm sgn}(m)\epsilon_{\mu\nu\rho}p_\rho/\sqrt{p_{0}}
\label{smatrix}
\end{equation}
(where $p_0$ is the energy) in the phase where there is mass generation $m$ for
the fermions.
{}From the breaking of the global $U(1)$ symmetry there is one Goldstone boson
but again there is no local order parameter. In that case the counting of the
inequality (\ref{uvir}) implies $ 1 + 3N > 2 $ i.e. $N > 1/3$. Note that in
this model, if one applies the Mermin-Wagner theorem, according to which there
is no well defined Goldstone boson, then the above inequality would only imply
$N \ge 0$. At any rate, the coupling of the system to 
electromagnetism, promotes the global fermion number $U(1)$ symmetry 
to a local electromagnetic symmetry, whose breaking 
results in superconductivity. In that case the would-be Goldstone boson
is eaten by the longitudinal component of the electromagnetic potential,
which now acquires a mass~\cite{Dorey:1990sz}.

\end{itemize}

This completes our discussion on the limitations of the 
applicability of 
the constraints of ref. \cite{appel} on
the three-dimensional gauge systems of interest to us here.
In the next section we proceed to 
review a SD analysis, 
which sheds some light on the non-asymptotic freedom of the 
effective QED$_3$ model, the absence of a critical number of 
fermion flavours for chiral symmetry breaking, but the existence
of a region of the effective charge where the phenomenon takes place. 
In addition we also discuss the non-trivial IR fixed point 
structure  of QED$_3$. We believe that 
all these features are physically important; especially
the last one implies the non-Fermi liquid behaviour of 
the relativistic spin liquid of nodal excitations under consideration.


\setcounter{equation}{0}
\section{SD equations and Critical Number of Flavours }

In this section we will present a study of the issue of mass generation and 
chiral symmetry breaking in the framework of the SD equations.

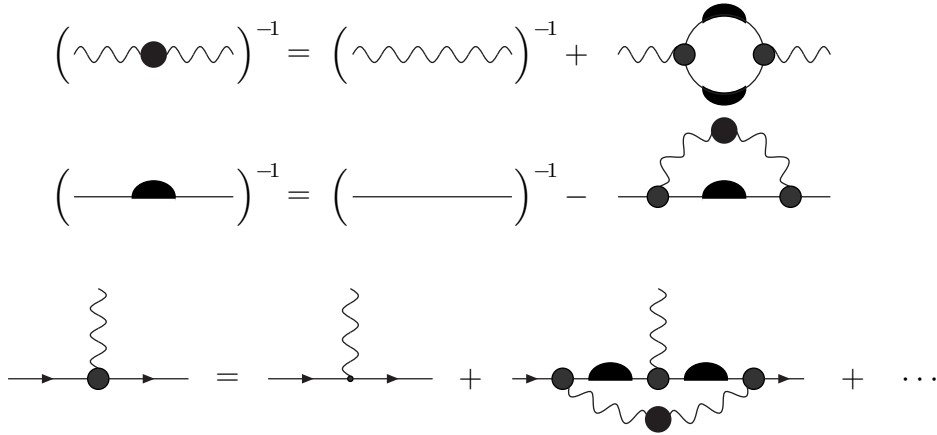
\begin{figure}[hbt]
\begin{center}
\begin{picture}(300,50)
\Text(5,30)[]{$\Big($}
\Photon(10,30)(35,30){3}{2.5}
\Vertex(40,30){5}
\Photon(45,30)(70,30){3}{2.5}
\Text(80,31)[]{$\Big)^{-\!1}$}
\Text(95,30)[]{$=$}
\Text(110,30)[]{$\Big($}
\Photon(115,30)(175,30){3}{5.5}
\Text(185,31)[]{$\Big)^{-\!1}$}
\Text(200,30)[]{$+$}
\Photon(215,30)(240,30){3}{2.5}
\Photon(270,30)(295,30){3}{2.5}
\GOval(255,43)(6,8)(0){0}
\GOval(255,17)(6,8)(0){0}
\CCirc(255,30){14}{White}{White}
\CArc(255,30)(15,0,360)
\GCirc(240,30){4}{0.2}
\GCirc(270,30){4}{0.2}
\end{picture}

\vspace{0.1cm}

\begin{picture}(300,50)(0,0)
\Text(5,30)[]{$\Big($}
\Line(10,30)(70,30)
\Text(80,31)[]{$\Big)^{-\! 1}$}
\GOval(40,30)(6,8)(0){0}
\CBox(32,23)(48,29){White}{White}
\Text(95,30)[]{$=$}
\Text(110,30)[]{$\Big($}
\Line(115,30)(175,30)
\Text(185,31)[]{$\Big)^{-\! 1}$}
\Text(200,30)[]{$-$}
\Line(215,30)(295,30)
\GOval(255,30)(6,8)(0){0}
\CBox(247,23)(263,29){White}{White}
\PhotonArc(255,30)(25,9,79){-3}{2.5}
\PhotonArc(255.4,30)(25,101,171){-3}{2.5}
\Vertex(255,55){5}
\GCirc(230,30){4}{0.2}
\GCirc(280,30){4}{0.2}
\end{picture}

\vspace{0.1cm}

\begin{picture}(330,65)(0,0)
\ArrowLine(0,30)(30,30)
\GCirc(34,30){4}{0.2}
\ArrowLine(38,30)(68,30)
\Photon(34,64)(34,34){3}{3}
\Text(83,30)[]{$=$}
\ArrowLine(98,30)(128,30)
\GCirc(129,30){1}{0.2}
\ArrowLine(130,30)(160,30)
\Photon(129,64)(129,31){3}{3}
\Text(175,30)[]{$+$}
\ArrowLine(190,30)(205,30)
\GCirc(209,30){4}{0.2}
\Line(213,30)(277,30)
\GCirc(245,30){4}{0.2}
\GCirc(281,30){4}{0.2}
\ArrowLine(285,30)(300,30)
\GOval(227,30)(6,8)(0){0}
\GOval(263,30)(6,8)(0){0}
\CBox(218,23)(236,29){White}{White}
\CBox(254,23)(272,29){White}{White}
\Photon(245,64)(245,34){3}{3}
\PhotonArc(245,64.7)(50,228,266){3}{3}
\PhotonArc(245,64.7)(50,274,312){-3}{3}
\Vertex(245,14.7){5}
\Text(315,30)[l]{$+\quad \cdots$}
\end{picture}
\end{center}
\caption{{\it The Schwinger-Dyson (SD) equations
for the photon and fermion self energies, and the vertex function.
The blobs indicate the
  full (non-perturbative) corrections.}}
\label{sdqed3}
\end{figure}

The derivation of the SD equations for the
photon propagator $\Delta_{\mu\nu}$, the electron
propagator $S_{F}$, and the photon-electron vertex
$\Gamma_{\mu}$ in QED$_3$ proceeds following standard methods
\cite{Cornwall:1974vz,cm}. The result is schematically 
depicted in fig. \ref{sdqed3}: 
\bea
\label{SD3}
\Delta_{\mu\nu}^{-1}(q) &=&
\Delta^{-1}_{0\mu\nu}(q)   +  e^2  \int  \frac{d^3k}{(2\pi)^3}
{\rm Tr}[\Gamma_{\mu} S_{F}\Gamma_{\nu}S_{F}] ~+\dots
\nonumber\\
S_{F}^{-1}(p) &=&  {S_{0F}^{-1}}(p)  - e^2 \int  \frac{d^3k}{(2\pi)^3}
\Gamma_{\mu}  S_{F}\Gamma_{\nu}\Delta^{\mu\nu}    ~+  \dots \\
\Gamma_{\mu}(p_1,p_2,p_3) &=&
\gamma_{\mu} -e^2  \int \frac{d^3k}{(2\pi)^3}
\Gamma_{\alpha} S_{F}\Gamma_{\mu}S_{F}
\Gamma_{\beta}\Delta^{\alpha\beta} +\dots \nonumber 
\eea
where $\Delta^{-1}_{0\mu\nu}(q) = q^2 g_{\mu\nu} + 
(\xi^{-1} -1) q_{\mu}q_{\nu}$, 
${S_{0F}^{-1}}(p)=\nd{p}-m$, and $p_1+p_2+p_3=0$.
The ellipses on the right-hand sides
denote the infinite set of terms containing the two-particle
irreducible
four-point function \cite{Cornwall:1974vz,cm}.
Although we are not working in the context of a large-$N$ analysis,
we note
that the above truncation is compatible with working to leading
order in resummed $1/N$ expansion.

We next define the scalar quantities
$A$, $B$ and ${\cal G}$ as follows:
\be
S_{F}(k)= \frac{1}{A(k)~\nd{k}},\,\,\,\,\,\,\,
\Delta _{\mu\nu}(k) = \frac{g_{\mu\nu}}{B(k)k^2}, \,\,\,\,\,\,\,
\Gamma _\mu (p_1,p_2,p_3)={\cal G}(p_1,p_2,p_3) \gamma _\mu
\label{photon}
\ee

Note that the form of Eq.(\ref{photon}) implies
that the longitudinal pieces of the photon propagator will
be discarded in what follows; this is motivated by the PT analysis
presented in the Appendix, 
particularly points {\bf{(iv)}} and {\bf{(v)}} 
of the last subsection.

\subsection{The semi-amputated vertex}

Following \cite{Cornwall:1989gv} and \cite{Papavassiliou:1991hx}
we define the semi-amputated vertex ${\hat G}$ as
\be
     {\hat G}(p_1,p_2,p_3) \equiv Z(p_1,p_2,p_3) {\cal G}(p_1,p_2,p_3)
\ee
\label{amp}
with
\be
Z(p_1,p_2,p_3) = B^{-1/2}(p_1)A^{-1/2}(p_2) A^{-1/2}(p_3)
\label{zed}
\ee
This definition proves very useful in
reducing the
complexity of the set of SD equations.
In addition, as explained in \cite{Mavromatos:1999jf}
the quantity
\be
g_R(p_1,p_2,p_3) \equiv e {\hat G}(p_1,p_2,p_3)
\label{runcoupl}
\ee
provides a
natural generalisation of the concept of the
running
or ``effective'' charge
in the 
context of super-renormalizable gauge
theories, such as QED$_3$.

The equation for the semi-amputated vertex $\hat{G}$
may be obtained from the third equation in (\ref{SD3}) 
by multiplying both sides by the
factor $Z(p_1,p_2,p_3)$, i.e.
\be
{\hat G}(p_1,p_2,p_3)\gamma_{\mu} =
Z(p_1,p_2,p_3)\gamma_{\mu}
-e^2  \int \frac{d^3k}{(2\pi)^3} {\hat G}^3
\gamma^{\alpha}\frac{1}{\nd{k}-\nd{p_1}}\gamma_{\mu} \frac{1}{\nd{k}}
\gamma_{\alpha}\frac{1}{(k+p_2)^2}
\label{SDamp}
\ee
where
${\hat G}^3 \equiv {\hat G}(p_3,k+p_2,p_1-k){\hat G}(p_1,-k,k-p_1)
{\hat G}(p_2,k,-k-p_2)$.
Restricting ourselves to the case where the photon
momentum is vanishingly small, one is left with a
single momentum scale $p$.
One can then define a
renormalization-group $\beta$ function from this ``running'' coupling
$G(p)$ by setting $\beta \equiv p \bigg(d{\hat G}(p)/d p\bigg)$.
In order to further simplify the SD equation for $G(p)$
we make the additional
approximation that ${\hat G}^3 = {\hat G}^3(k)$, i.e. a cubic power
of a single ${\hat G}(k)$
depending
only on  the integration variable $k$.

Carrying out the
$\gamma$-matrix algebra in $d=3$-dimensional Euclidean space,
one obtains:
\be
{\hat G}(p) =
Z(p) + \frac{1}{3}e^2 \int \frac{d^3k}{(2\pi)^3} {\hat G}^3 (k)
\frac{1}{k^2 (k-p)^2}
\label{asymptotic}
\ee

We observe that $Z(p) \rightarrow 1$ for
$p \rightarrow \infty$, where perturbation theory is valid.
This is expected
from the fact that in such a case
the functions  $A(p), B(p) \rightarrow 1$ trivially.
Moreover, from (\ref{asymptotic}) we see that,
if ${\hat G}$ stays positive, which is expected for any physical
theory, then,
as a result of the positivity of
the integrand, $G(p) \ge 1$ for any $p$.
Thus, one has the following basic properties of ${\hat G}(p)$,
which stem directly from the integral equation (\ref{asymptotic}):
${\hat G}(p) \ge 1$, ({for~all}~ $p$), and
${\hat G}(p) \rightarrow 1, \quad p \rightarrow \infty$.
Assuming that in the IR regime,
 $k/\alpha << 1$,
the inhomogeneous term
$Z(p)$
goes
to zero,
(this assumption has been justified by explicit
calculations in \cite{Mavromatos:1999jf})
one can decouple
the equation for the {\it amputated} vertex
from the equations for $A(p)$, $B(p)$. Thus, one
arrives at the {\it homogeneous} integral equation
\be
     {\hat G}(p) =
\frac{1}{3}e^2 \int \frac{d^3k}{(2\pi)^3} {\hat G}^3 (k)
\frac{1}{k^2 (k-p)^2}
\label{ampapp}
\ee
involving only one unknown function, namely ${\hat G}$,
which must be self-consistently determined.
Note that Eq.(\ref{ampapp})is invariant under the rescaling
${\hat G} \rightarrow {\hat G}/e$. This indicates
a straightforward extension of the analysis to a large-$N$ treatment,
given that $N$ appears only as a multiplicative factor.
Eq.(\ref{ampapp}) does not admit {\it physically acceptable} solutions, i.e.
solutions with ${\hat G} \ge 0$ and {\it finite}.
Indeed, setting
$p =0$ one obtains after the (trivial) angular integration

\be
       {\hat G}(0) = \frac{e^2}{12\pi^2} \int _0^\infty \frac{dk}{k^2}
{\hat G}^3(k)
\label{limito}
\ee
Finiteness of ${\hat G}(0)$ requires that the integrand of the right hand side
of (\ref{limito}) converges at $y \rightarrow 0~{\rm and} ~\infty$.
The UV limit does not present a problem, because the
kernel vanishes like $y^{-2}$, which is consistent with the
super-renormalizability of the theory as well as the fact that
the amputated vertex tends to 1.
In the IR limit $y \rightarrow 0$, however,
the kernel blows up.
For the integral to remain finite at that point, as required
by the finiteness assumption for ${\hat G}(0)$,
$G^3(y)$ must approach zero as $y^\alpha,~\alpha > 1/3$, thereby
implying that ${\hat G}(0) =0$.
However for that to happen the integrand in (\ref{limito})
must change sign, which would in turn imply that ${\hat G}(y)$ itself
must change sign somewhere in $y$.
According to our assumption above this
is not a physically acceptable situation.

The way to see that
indeed the behaviour $G(y) \sim y^{1/2}$ as $y \rightarrow 0$,
would be the {\it only } possibility is to
convert
the integral equation into a non-linear differential equation.
To this end, we perform the angular integration in (\ref{ampapp}),
to arrive at the equation:
\be
  {\hat G}(p) = \frac{2}{3\pi^2} \frac{\alpha}{p} \int _0^\infty \frac{dk}{k}
{\hat G}^3(k) {\rm ln}|\frac{k+p}{k-p}|
\label{v2}
\ee
where we have set $e^2 \equiv 8\alpha$ to make contact with the
usual large-$N$ definition~\cite{app}. For us, however,
the number of fermion flavours is not assumed to be  necessarily
large.
Introducing the dimensionless variables
$x \equiv p/\alpha$ and $y\equiv k/\alpha$, one obtains
in the limit $x <<1 $
\be
{\hat G}(x) =\frac{2}{3\pi^2 x } \int _0^\infty
\frac{dy}{y} {\hat G}^3 (y)
{\rm ln}|\frac{y+x}{y-x}| \simeq
\frac{4}{3\pi^2} \Bigg(
\frac{1}{x^2}\int _0^x dy {\hat G}^3(y) +
\int _x^\infty \frac{dy}{y^2} {\hat G}^3 (y)\Bigg)
\label{v4}
\ee
Differentiating appropriately with respect to
$x$, and conveniently rescaling ${\hat G}$ by setting
$G \equiv \sqrt{\frac{8}{3\pi^2}} {\hat G}$,
we arrive at the following differential equation for small $x$:
\be
x^3 \frac{d^2 G}{d x^2} + 3 x^2 \frac{d G}{d x} +
G^3(x) =0 , \qquad x << 1
\label{fcaleq2}
\ee
The obvious solution of this equation, for $x << 1$, is
$G(x) = \frac{\sqrt{5}}{2}x^{1/2}$.
Note that this
solution would imply a `trivial  IR fixed point structure'
given that its associated $\beta$ function vanishes at $x=0$.
As we shall demonstrate below this is in fact the only solution
with $G \ge 0$. Indeed,
upon the change of variables
$\xi^{-1} = 2 x^2$, $G=2^{3/4}~\eta(\xi)$,
the equation becomes of {\it Emden-Fowler} type~\cite{emden,kamke}:
\be
    \frac{d^2}{d \xi^2} \eta (\xi) - \xi^{-3/2} \eta^3 (\xi) =0, \qquad
\xi \rightarrow +\infty
\label{emden2}
\ee
As discussed in the mathematical literature~\cite{emden},
the {\it only positive solution}  of (\ref{emden2}), as $\xi \rightarrow +\infty$,
has the form as $\xi \rightarrow +\infty$:
\be
      \eta (\xi) = \frac{\sqrt{5}}{4} \xi^{-1/4}, \qquad
\label{onlysol}
\ee
which is exactly the solution $G(x) = \frac{\sqrt{5}}{2}x^{1/2}$ found above,
in the region $x \rightarrow 0$.

The above analysis suggests that
no non-trivial IR-fixed point is possible in QED$_3$ in the absence
of an IR cut-off, as already conjectured  in ref. \cite{aitch}.

\subsection{The IR fixed point}
We next study the behaviour of the SD equation for ${\hat G}(p)$ in the
presence of an IR  cut-off.
W shall consider the case of a fermion
mass gap $m(p)= \Sigma (p)/A(p)$, where $\Sigma (p)$ is
the fermion self energy.
In that case the fermion propagator $S_F^{-1}$ becomes:
\be
     S_F (k) = \frac{1}{A(k) \left( \nd{k} + m_f(k)\right)}
\label{fermass}
\ee
 and we assume that $m_f(p) \simeq m_f(0) \equiv m_f \ne 0$.
In that case the integral equation (\ref{ampapp}) becomes:
\be
   {\hat G}(p) =
\frac{e^2}{3} \Bigg( \int \frac{d^3k}{(2\pi)^3}
\frac{{\hat G}^3 (k)}{(k^2+ m_f^2 )(k-p)^2} + 2m_f^2
\int \frac{d^3k}{(2\pi)^3}
\frac{{\hat G}^3 (k)}{(k^2 + m_f^2)^2 (k-p)^2} \Bigg)
\label{ineq2}
\ee
Performing the angular integrations
one arrives at:
\be
{\hat G}(x) = \frac{2}{3\pi^2 x}\int dy f(y)
{\rm ln}\left|\frac{y+x}{y-x}\right|{\hat G}^3(y)
\label{equntitl}
\ee
where $x \equiv p/\alpha$, $m \equiv m_f/\alpha$ are dimensionless,
and
\be
  f(y) \equiv y \frac{y^2 + 3m^2}{(y^2 + m^2)^2} \ge 0
\label{fdef}
\ee
Differentiation with respect to $x$ yields:
\be
 x\frac{d}{d x}{\hat G}(x) =-\frac{2}{3\pi^2 x}\int _0^\infty
dy f(y) \left( {\rm ln}\left|\frac{y+x}{y-x}\right|+\frac{2xy}{x^2-y^2}\right)
{\hat G}^3(y)
\label{kernel}
\ee
One observes that formally as $x\rightarrow 0$ the right-hand-side vanishes,
provided that ${\hat G}$ is finite. This indicates the existence
of a fixed point. As we shall show below this is confirmed
analytically by converting the integral equation into
a non-linear differential equation.
To accomplish this,
one expands the logarithms
for small $x << 1$, and then differentiates with respect to $x$, arriving at
\be
x(x^2 + m^2)^2 \frac{d^2}{d x^2}{\hat G}(x) +
3(x^2 + m^2)^2\frac{d}{dx}{\hat G}(x) +
\frac{8}{3\pi^2}(x^2 + 3m^2){\hat G}^3(x) =0
\label{emdenmass}
\ee
In the IR region
$x << m$ the equation (\ref{emdenmass}) is approximated by:
\be
    x \frac{d^2}{dx^2}{\hat G}(x) + 3 \frac{d}{d x}{\hat G}(x) +
\frac{8}{\pi^2m^2}{\hat G}^3(x)=0
\label{papmav}
\ee
and is immediate to see that
a special power-law solution is given by (for positive $G(x)$)
by:
\be
 {\hat G}(x) =m \pi \frac{\sqrt{3}}{4\sqrt{2}}x^{-1/2}
\label{ss}
\ee
Notice the IR divergence of this type of solutions {\it even}
in the presence of a (bare) fermion mass.
The associated renormalization-group $\beta$ function
for this case reads:
\be
     \beta (x) = -\frac{1}{2}{\hat G} \sim x^{-1/2}
\rightarrow +\infty,~{\rm as}~x \rightarrow 0
\label{beta2}
\ee
indicating the absence of an IR fixed point.
The associated operator appears to be {\it relevant}
(negative scaling dimension), which implies the possibility of the theory
driven to a non-trivial fixed point.

However, in the IR regime $x << 1$,
one can find a different type of solution:
\be
    {\hat  G} =  m \pi \frac{\sqrt{3}}{2\sqrt{2}}\frac{c}{1 + c^2 x}, \qquad x \rightarrow 0
\label{es}
\ee
where $c$ is a constant of integration to be fixed
by the boundary condition at $x=0$ implied by the integral equation,
to be discussed later on. For physical solutions $c$ is assumed positive.
This type of solutions has a renormalization-group $\beta$-function
of the form:
\be
     \beta = -{\hat G}(x) + \frac{2\sqrt{2}}{\sqrt{3}\pi m c}{\hat G}^2(x) \sim
-\frac{x}{\left(1 + c^2 x\right)^2} \rightarrow 0, \qquad x \rightarrow 0
\label{beta3}
\ee
from which we observe the existence of a non-trivial (non-perturbative) IR fixed
point at ${\hat G}^* = \frac{\pi m \sqrt{3}c}{2\sqrt{2}} >0$.
Such a fixed point
is the result of the dynamical generation of a
parity-invariant, chiral-symmetry breaking
fermion mass~\cite{app}, indicating
the connection of the phenomenon of chiral symmetry breaking
in QED$_3$ with a non-trivial IR fixed point structure.

The non-trivial fixed-point solution (\ref{es}) 
is compatible with the integral equation (\ref{ineq2}) 
for {\it some } values of the fermion mass $m$ to be specified below. 
Indeed, one can derive a 
{\it boundary condition} for ${\hat G}(0)$ from (\ref{ineq2}), which reads:
\be
 {\hat G}(0) 
= \frac{4}{3\pi^2}\int _0^\infty dy \frac{{\hat G}^3(y)}{y^2 + m^2}  
+ \frac{8m^2}{3\pi^2}\int _0^\infty dy \frac{{\hat G}^3(y)}{(y^2 + m^2)^2}
\label{bc2}
\ee
In contrast to the massless case (\ref{limito}), ${\hat G}(0)$ 
is now a finite constant, $\pi m c \sqrt{3/8}$, as seen from (\ref{es}), 
and this allows for a compatibility of the solution (\ref{es}) 
with (\ref{bc2}), {\it provided} that $m {\hat G}(0)$ 
satisfies certain conditions. Such conditions have been derived 
in~\cite{Mavromatos:1999jf}, where a full analysis of the 
\emph{non linear} coupled system 
of vertex, fermion and photon self-energy integral equations
has been performed. The result is summarized in the equation: 
\be
   m < \frac{\frac{8}{3\pi}\left(1 - \frac{1}{\pi}\right)}
{{\hat G}(0) - \frac{12\sqrt{6}}{5}}, \qquad {\hat G}(0) > 
\frac{12\sqrt{6}}{5} \simeq 5.88~.
\label{fincond}
\ee
This condition will act as a boundary condition for the 
allowed values of $m$ in a mass-coupling 
diagram, as we shall discuss later on.

\subsection{Dynamical generation of the fermion mass gap}
In the previous subsection we have assumed
the presence of
a finite fermion mass,
which we have treated effectively as
an arbitrary parameter of the model.
In this subsection we turn to the full
problem, and study the dynamical generation
of this mass, by deriving it self-consistently
from the corresponding SD mass-gap equation.

The equation for the gap $\Sigma (p)$ reads \be
A(p)~\nd{p} + \Sigma (p) = \nd{p} + A(p)e^2 \int \frac{d^3k}{(2\pi)^3}
{\hat G}^2(k) \gamma _\mu \frac{1}{\nd{k} + M(k)}\gamma ^\mu \frac{1}{(k-p)^2}
\label{massgap}
\ee
where $M(k) \equiv \Sigma (k)/A(k)$ is the mass function, and
we have pulled out factors of $A(p)$ appropriately so as to be able
to define an amputated vertex function ${\cal G}(k)$.
After standard algebraic
manipulations, using dimensionless variables, in units of $\alpha=e^2/8$,
${\tilde M}\equiv M(k)/\alpha$, $x\equiv p/\alpha$, $y \equiv k/\alpha$,
and working in the regime of low momenta $x << 1$,
one arrives at the following differential equation:
\be
x \frac{d^2}{d x^2} {\tilde M}(x) + 3 \frac{d}{dx}{\tilde M}(x)
+ \frac{24}{\pi^2}\frac{{\tilde M}(x)}{x^2 + {\tilde M}^2(x)}{\hat G}^2(x) = 0
\label{gapeq}
\ee
In the relevant region
$x^2 << {\tilde M}^2 << 1$,
we neglect $x^2$ next to ${\tilde M}^2$ in (\ref{gapeq})
and
use the solution (\ref{es}) for ${\hat G}(x) \simeq
{\tilde M}\sqrt{\frac{3}{8}}\pi c$ as $x \rightarrow 0$. The result is:
\be
x \frac{d^2}{d x^2} {\tilde M}(x) + 3 \frac{d}{dx}{\tilde M}(x)
+ 9c^2{\tilde M}(x)= 0, \qquad x \rightarrow 0
\label{bessel}
\ee
from which one obtains a power series expression for the dynamical mass:
\be
  {\tilde M}(x) = C_1 x^{-1} \sum _{n=0}^{\infty} (-1)^n
\left(3c\right)^{2n + 2} x^{1 + n} \frac{1}{ n! \Gamma (3 + n)}
\simeq \frac{9}{2} C_1 c^2 + {\cal O}(x), \qquad x \rightarrow 0
\label{fermionmass}
\ee
{}From this one obtains the following relation between ${\hat G}(0)$ and
${\tilde M}(0)\equiv m_f /\alpha $:
\be
       {\tilde M}(0) \equiv m_f/\alpha  \equiv {\tilde m} = 
= \left(\frac{12}{\pi^2}\right)^{1/3}C_1^{1/3}{\hat G}^{2/3}(0)
\label{appelgm}
\ee

\subsection{Comparison with the large $N$}

It is important to compare the above results with those
obtained within the context of a large-$N$ analysis~\cite{app}.
In particular, at first sight it seems that the relation
(\ref{appelgm}) does not have a critical coupling, above which
dynamical mass generation occurs.
However, because the result (\ref{appelgm}) has been derived
in the context of the solution (\ref{es}),
one should bear in mind the restrictions characterizing this
situation.
In particular, as has been explained in detail in  \cite{Mavromatos:1999jf}
the following conditions must hold:
\be
{\hat G}^{5/3}(0) -
\frac{12\sqrt{6}}{5}{\hat G}^{2/3}(0) -
\frac{8}{3 \left(12 \pi C_1\right)^{1/3}}\left(1 -
\frac{1}{\pi}\right) < 0 , \qquad
{\hat G}(0) > \frac{12\sqrt{6}}{5}
\label{c1}
\ee
This restriction implies a critical coupling,
${\hat G}_c = 12\sqrt{6}/5 \simeq 5.88$
but it is derived
in a way independent of any large-$N$ analysis.
The way to understand (\ref{c1}) is the following: one should first
fix a range of ${\hat G}(0)$, with ${\hat G}(0) > 5.88$, and then
use a $C_1$ that will be such that, within this range of the couplings,
eq. (\ref{c1}) is satisfied for masses ${\tilde m} << 1$.
As can be readily seen, the bound for $C_1$ obtained from the
requirement
that $m <<1$ is far less restrictive than the one
associated with (\ref{c1}), provided
${\hat G}(0)$ is not too close to the critical ${\hat G}_c(0)$,
where the mass $m$ vanishes of course.
For instance, for ${\hat G}(0) ={\cal O}(8)$, the upper bound
on $C_1 $ from (\ref{c1}) is of order ${\cal O}(10^{-4})$, while
for ${\hat G}(0) = 6$ the upper bound is $C_1 < 4$. Notice that
the bound is
very sensitive to small changes in ${\hat G}(0)$.

\begin{figure}[htb]
\epsfxsize=3in
\bigskip
\centerline{\epsffile{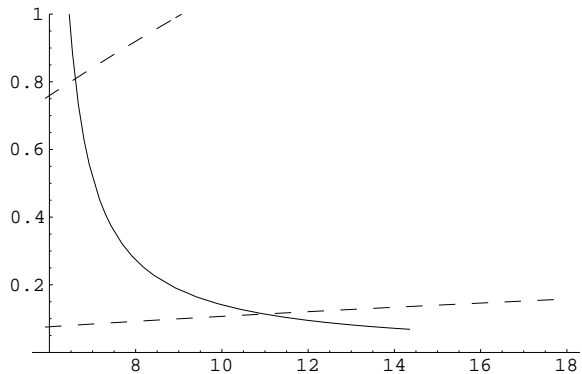}}
\caption{\it\baselineskip=12pt 
Fermion mass versus the IR-value 
of the coupling ${\hat G}(0)$
using (\ref{appelgm}) (dashed curves), for two values of 
$C_1=10^{-5}$ (lower dashed curve) and $C_1=10^{-2}$ (upper dashed curve). 
The continuous curve is (\ref{fincond}), viewed as a boundary condition.
The value $C_1 =10^{-2}$ should be excluded on grounds of yielding 
too high values of the mass ${\tilde m}$.}  
\bigskip
\label{gmpm}\end{figure}

A typical situation is depicted in fig. \ref{gmpm} for two values
of $C_1 =10^{-5}, 10^{-2}$. We observe that the case $C_1=10^{-2}$ yields
an upper bound in the mass which is of order $0.8$ and hence
should be discarded on the basis that it is not small enough.
On the other hand the value $C_1=10^{-5}$ yields an acceptable upper bound
$ m \sim 0.1$. In that case, from fig. \ref{gmpm}, we observe that
the allowed region of $m$ is
\be
0.08 \lsim m \lsim 0.12
\label{zone4}
\ee
The corresponding regime of the couplings ${\hat G}(0)$
is:
\be
   5.88 < {\hat G}(0) < 11
\label{regionc4}
\ee
The above results are to be compared with 
the corresponding regimes of masses and couplings derived 
in a large-$N$ analysis. 
We recall that, in the context of a large $N$ treatment, and 
to leading order in $1/N$ resummation, the following solution
for the dynamically-generated $m$  
is found~\cite{app} 
\be
    m \sim {\cal O}(1) {\rm exp}\left(-\frac{2\pi}{\sqrt{\frac{g^2}{g_c^2}-1}}\right)
\label{appelquist}
\ee
where $g_c^2 =\frac{\pi^2}{32}$ is the critical coupling, above which
dynamical mass generation occurs~\cite{app}.
In fact $g_c$ is interpreted as an inverse critical number $1/N_c$ 
of four component 
fermion flavours. When $1/N^2$ or higher-order 
corrections are included, 
the dynamical mass 
generation procedure 
yields $3 < N_c < 4$ (a result that as we have seen has been refuted 
by the inequality of~\cite{appel}, discussed in the 
previous section). 

\begin{figure}[htb]
\epsfxsize=3in
\bigskip
\centerline{\epsffile{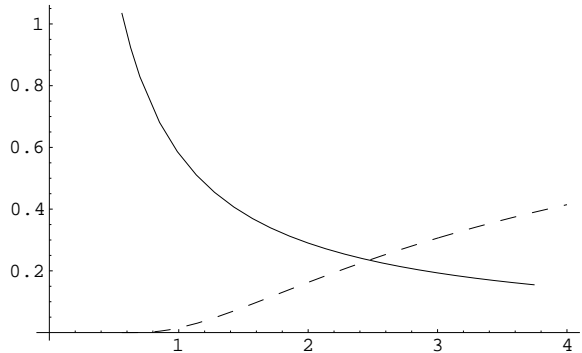}}
\caption{\it\baselineskip=12pt 
Fermion mass versus the IR-value 
of the coupling ${\hat G}(0)$. The solid curve 
represents the condition 
derived from the integral equation for the vertex, 
whereas the dashed 
line represents the solution obtained from 
the standard gap equation in the 
large-$N$ treatment.}
\bigskip
\label{fig3}\end{figure}

Compatibility of the solution (\ref{appelquist}) 
with the constraints obtained from the integral equations for the vertex 
implies~\cite{Mavromatos:1999jf} 
the existence of an {\it upper bound} on fermion masses,  
$m < m_{max}$, where $m_{max}$ is defined
through the intersection of appropriate 
curves coming from the non linear constraints (c.f. fig. \ref{fig3}).
This yields $m_{max} \simeq 0.3$.

On the other hand, for large momenta, we know that ${\hat G} \rightarrow 1$.
Physically one expects
{\it a monotonic decrease}  of ${\hat G}(x)$ over the {\it entire}
range of $x \in [0, \infty)$. This would occur in our case 
if and only if ${\hat G}(0) > 1$, which,
in the context of the large-$N$ result of \cite{app,Dorey:1990sz}, implies
a {\it minimum} bound for the fermion masses   
$m > m_{min}\simeq 0.03$. Actually, as we shall argue 
in the next section, ${\hat G}(0)$ should be comfortably 
larger than $\sqrt{3/2}$ 
for self-consistency of our approximations. 

Hence, we see that the monotonicity of the
running coupling can be achieved 
in the context of a large-$N$ treatment, 
if 
the mass $m$ lies in the following regime:
\be 
      0.03 \lsim m \lsim 0.3 
\label{zone}
\ee
or equivalently if the coupling at the IR 
point ${\hat G}(0)$ is restricted in the regime:
\be
           1 < {\hat G}(0) < 2.5,\qquad 
\label{regionc}
\ee
which is to be compared with (\ref{regionc4}). 

{}From the physical point of view
of applicability of the 
QED$_3$ models to the theory of high temperature 
superconductors, 
the above results imply constraints on the 
microscopic parameters of the underlying lattice models,
whose long-wavelength limit is the three dimensional 
gauge theory under consideration. 
It must be noted in this case that 
the above analysis applies equally well to both QED$_3$ 
and $\tau_3$-QED models.
 For instance, in the models of~\cite{farakos,Dorey:1990sz}
the effective (gauge-invariant) 
coupling ${\hat G}(0)$ may be expressed  
in terms of the parameters 
of the microscopic condensed-matter lattice systems, 
as:
\be
    {\hat G}(0)^2 \sim \frac{J}{e^2} (1-\delta)
\label{dopheis}
\ee
where 
$\delta$ expresses
the concentration of impurities in the system (doping), and 
$J$ denotes the Heisenberg (antiferromagnetic) exchange energy. 
Hence, on account of 
(\ref{regionc4}), Eq. (\ref{dopheis}) implies 
that $ 6 \lsim (J/e^2)(1 - \delta) \lsim 11$ for superconductivity 
to occur. 
In 
phenomenologically acceptable models~\cite{Dorey:1990sz}
$e^2/J \sim 0.1$, which implies 
an upper bound on $\delta \sim 0.4$, which is physically 
consistent. By slightly modifying this ratio we may even obtain 
also lower limits for $\delta$, and hence a region where the 
phenomenon of fermion mass generation of the nodal excitations occurs.  
However, the reader should bear 
in mind that the above-described limiting values
are rather indicative at present, given 
that a complete quantitative understanding of the 
underlying dynamics of high-temperature superconductivity
from an effective gauge-theory point of view  
is still lacking. 

\setcounter{equation}{0}
\section{Conclusions and Outlook}

In this review we have described some unconventional phases of 
$QED_3$ and related models, occurring at $T=0$, including KT superconductivity.
Such features of QED$_3$ are important when discussing the 
phase diagrams of underdoped AF. For the novel phases to occur it is necessary 
to have a dynamical generation of fermion mass for the case of 
{\it two} four-component spinors, which corresponds to the physical case. 

This last feature 
has been questioned recently in \cite{appel}, by resorting to 
a conjectured inequality. Should this inequality 
be applicable in this context, it would strongly disfavour 
the generation of chiral symmetry breaking fermion mass for the case {\it two} 
four component spinors, contrary to the findings of various analysis
 based on SD equations.
However, we have presented a number of arguments as to why, in our opinion, 
the conditions necessary for the applicability of the inequality in question
\cite{appel} are not fulfilled in the relevant models.

One might expect 
that a future resolution of some of the above issues 
can be furnished by lattice models: 
QED$_3$ with an even number of fermion flavours 
can indeed be simulated on the lattice~\cite{hands}, as it 
does not suffer from a fermion determinant sign problem. 
However, this is {\it not the case}
for the non-compact 
$\tau_3$-QED~\footnote{We thank S. Hands for 
an informative discussion on this point.},
whose simulations on a lattice still remain a big challenge, due to 
a sign problem in the 
respective fermion determinant,
as a result of the $\tau_3$ structure of the statistical gauge fields.
Nevertheless, the compact case of $\tau_3$ QED, which can be 
considered as a broken phase of a SU(2) gauge theory, can be simulated. 

However, as remarked, in \cite{temp}, 
caution should be exercised 
when one simulates QED$_3$ models on the lattice, due to  
the IR cutoff-dependence 
of many of the quantities entering the pertinent 
chiral symmetry breaking dynamics, and in particular the 
critical number of fermion flavours. 
This is because of the non-trivial 
IR divergences of the model, which need regularization by means of the 
scale $\mu$~\footnote{It is worth remarking at this 
stage that the dependence of 
$N_c$ on the IR cutoff mass scale, $\mu$, is actually similar
to its temperature dependence, discussed in \cite{aitch}, given that,
under certain circumstances,
one may associate the temperature with an IR cutoff.}. 

In \cite{temp} both a numerical study and an analytic (approximate)
expression have been given for the dependence of $N_c$ on $\mu$:
$(\nu_c/2) {\rm ln}(\alpha/\mu) + 2{\rm arctan}(\nu_c) = \pi$,
with $\nu_c =\sqrt{32/(N_c\pi^2) - 1}$, and $\alpha = Ne^2/8$, and $e$ the 
QED$_3$ bare coupling (with dimensions of square root of mass). 
>From either the above analytic formula or the numerical analysis 
presented in \cite{temp} follows that,  
in order to find 
chiral symmetry breaking for $N=2$, one should use {\it at least}
a ratio $\pi e^2/\mu = 5 \times 10^3$. 

Note that in order to extract safe conclusions 
for chiral symmetry breaking on a lattice, 
with a dimensionless 
lattice coupling $\beta \simeq 1/e^2a$, where $a$ is the lattice spacing, 
it is imperative that the volume of the lattice $L^3$ should be large
compared to any other dynamically generated correlations of the system. 
Thus, the physical volume $(L/\beta)^3$ must be large.
If we set $La \sim \pi/\mu$~\cite{temp},
then $L/\beta \sim \pi e^2/\mu$,
and hence to observe a chiral symmetry breaking for the case of 
two four-component fermion flavours ($N=2$) one needs $L/\beta > 10^3$, 
according to the 
above remarks. 
However, 
typical lattice simulations in the existing literature~\cite{hands} 
use $L/\beta = {\cal O}(10^2)$. This, in turn, may explain 
why chiral symmetry breaking is not seen in such cases (for $N=2$). 

Given the practical difficulties in reaching such high values 
of $L/\beta ={\cal O}(5 \times 10^3)$ 
(with the presently available computing power), 
one may want to explore the possibility of
extracting exact results on some 
of the physical issues regarding the novel quantum critical 
phases discussed in the 
present article. One such possibility would be to 
embed the gauge theories in question into some 
{\it supersymmetric} (SUSY) theory, with at least ${\cal N}=2$ conserved 
SUSY charges, which is the minimum number of supersymmetries
to allow for exact results in (2+1)-dimensions.
This has been argued to be possible in the 
context of semi-realistic condensed matter situations in \cite{ams},
from the point of view of a {\it composite operator} field theory, 
based on the dynamics of composites consisting of spinons and holons. 
The basic idea is that at {\it certain} regions of the parameter 
space of appropriate microscopic condensed matter models (such as extended
$t-j$ models with next-to-nearest neighbor interactions~\cite{sm}),
there is an effective supersymmetry between spinons and holons,
characterising the dynamics of the effective continuum field theory
near nodal points. This
symmetry, originally an ${\cal N}=1$ SUSY, can be elevated to an extended 
${\cal N}=2$ SUSY, as a result of the low-dimensionality of the 
model, implying the existence of a {\it topologically conserved} current,
which supplies the latter supersymmetry structure~\cite{ams}. 

The reason for working with composites, instead 
directly with spinon and holon 
constituents, lies on the fact that in this way one obtains dynamically
massless gauge fields (Abelian), as particular combinations 
of spinons and holons. The resulting ${\cal N}=2$ SUSY 
transformations of the composite fields have been ensured 
up to quartic order in the spinon and holon constituents in \cite{ams},
where we refer the interested reader for details. In this approach 
only the spinon (magnon) bosonic fields, $z$, and the fermionic holon 
fields, $\psi$, are fundamental degrees of freedom. 
The {\it statistical} gauge fields
are {\it induced} as particular combinations of these fields, consistent 
with a composite ${\cal N}=2$ SUSY at specific regions of the 
parameter space of the microscopic condensed-matter model. 

The resulting effective continuum model turns out to be the 
${\cal N}=2$ SUSY Abelian Higgs model. There are some exact results
associated with the phase structure of the model, in particular
the {\it topology} of the so-called moduli space, i.e. the 
space of the (complex) parameters of the SUSY model. These results
are detailed in \cite{ams} and we shall not discuss them here. 
However, we do note that they include a passage from a pseudogap
to an unconventional KT superconducting phase for the composite model,
in the compact gauge field case, 
of similar nature to the one discussed here for the constituent
excitations of spinon and holons. 
This passage may be 
understood qualitatively as follows:  
the ${\cal N}=2$ SUSY Abelian Higgs model 
(as is otherwise called the ${\cal N}=2$ SUSY QED$_3$ (SQED))
possesses a Higgs phase, where the gauge field is massive; this phase  
has been argued to correspond to 
a pseudogap phase, which, in the compact SQED case, 
may also be characterised by stripe-like
configurations, due to domain walls of the compact gauge field~\cite{ams}. 
The compact SQED contains monopoles with {\it both} negative and positive 
charges, and the corresponding antimonopoles. 
The fugacities of the monopole configurations of the compact 
case may depend on doping concentration in such a way that 
at certain doping concentrations the fugacities of the negative charge 
monopoles {\it vanish}, leaving only monopoles of positive charge (say +1),
and antimonopoles of charge -1. Such a case is similar to the 
case where the compact U(1) gauge theory is embedded into an SU(2) 
gauge theory.
In such SU(2)-like theories it can be shown~\cite{ams} that 
there are no
stripe phases, and moreover the statistical photon remains 
exactly massless~\cite{ahw}, thereby implying the onset of a KT superconducting
phase, as in \cite{Dorey:1990sz}, 
and the end of a striped pseudogapped phase. 

Finally, in \cite{Campbell-Smith:1999aw} the issue of dynamical mass generation 
in ${\cal N}=1$ SUSY QED$_3$ was studied using superfield SD equations. It was 
shown that the presence of a supersymmetry-preserving mass for the matter
multiplet stabilizes the infrared gauge couplings against oscillations present in 
the massless case, thus inferring that the massive vacuum is dynamically
selected at the level of the quantum effective action.

It must be noted that such supersymemtric points in general do not correspond
to realistic values of the parameters of the underlying condensed-matter
microscopic model. Nevertheless, they may be viewed as implying 
the possibility of using such extended supersymmetry 
(between spinon and holon composites) as a tool for extracting
exact analytic information on the phase diagram of doped antiferromagnets
as follows: one studies first such extended SUSY points, 
and then breaks the ${\cal N}=2$ SUSY 
explicitly down to ${\cal N}=0$ by varying the 
parameters of the model in order to approach the physical regime. 
One may then hope to construct models in which such a breaking 
would result in a situation where 
the SUSY partners of the physical excitations acquire masses higher than
the highest mass scale in the problem (e.g. Debye screening), and hence 
can be safely discarded. Such issues definitely deserve further studies, 
and indeed may pave the way for obtaining  an exact analytic understanding
of the complicated phase diagram of doped antiferromagnets, without 
resorting to lattice simulations of the models. It remains to be seen 
whether such hopes can be realised within the context of 
phenomenologically realistic condensed-matter theories.

\section*{Acknowledgments}

We thank J.Alexandre, S. Hands, P.A. Marchetti, Sarben Sarkar and P. Sodano
for informative discussions. This work has been developed 
during the workshop 
{\it Hidden Symmetries in Strongly Correlated Electron Systems and Low 
Dimensional Field Theories}, 
organised by the Physics Department of King's 
College London, July 8-12 2003. The work of N.E.M. is partly 
supported by a Visiting Professorship at the University of Valencia,
Department of Theoretical Physics, the Leverhulme Trust (U.K.) and 
the European Union (contract HPRN-CT-2000-00152). 
The work of J.P. is supported by the CICYT Grant FPA 2002-00612.

\setcounter{equation}{0}
\section*{Appendix: The Pinch Technique}

It is well known that off-shell Green's functions 
depend in general on
the  gauge-fixing  procedure  used  to  quantize the  theory,  and  in
particular on  the gauge-fixing parameter (GFP) chosen  within a given
scheme. The   fermion  self-energy  $\Sigma  (p)$, for example, is
GFP-dependent already  at the one-loop  level.  
The dependence  on the
GFP is  in general non-trivial and  affects the properties  of a given
Green's  function. In QED$_4$, and  
in the  framework  of  the  covariant gauges,
depending on  the choice of the GFP  $\xi$, one may eliminate
the UV divergence of the one-loop electron propagator $\Sigma
(p,\xi)$  by  choosing  the  Landau  gauge $\xi=0$,  or  the  IR
divergence  appearing after on-shell  renormalization by  choosing the
Yennie-Fried   gauge  $\xi=3$.   The   situation  becomes   even  more
complicated  in the  case  of non-Abelian  gauge  theories, where  all
Green's  functions  depend  on  the  GFP.   Of  course,  when  forming
observables  the gauge  dependences  of the  Green's functions  cancel
among  each  other order  by  order  in  perturbation theory,  due  to
powerful  field-theoretical  properties, a  fact  which reduces  their
seriousness.  However, these dependences  pose a major difficulty when
one  attempts  to   extract  physically  meaningful  information  from
individual  Green's functions.   This  is  the case when studying
the SD equations; this infinite
system of  coupled non-linear integral 
equations  for all Green's  functions of
the  theory is inherently  non-perturbative  and can
accommodate phenomena  such as  chiral symmetry breaking  and dynamical
mass  generation.   
In  practice one  is  severely
limited  in their  use,  and a  self-consistent  truncation scheme  is
needed. The main  problem in this context is  that the SD
equations are  built out  of gauge-dependent Green's  functions; since
the  cancellation  mechanism  is  very subtle,  involving  a  delicate
conspiracy of terms  from {\it all orders}, a  casual truncation often
gives   rise   to   gauge-dependent  approximations   for   ostensibly
gauge-independent quantities \cite{Cornwall:1974vz,Marciano:su}.
 The study of SD
equations, and especially of ``gap equations'', 
has been  particularly popular in QED \cite{Johnson:1964da},
and  even  more  so  in  QCD \cite{Lane:1974he},  where  it  has  been
intimately  associated  with  the  mechanism that  breaks  the  chiral
symmetry.   Similar  equations  are  relevant in  QED$_3$,  where  the
IR regime of the theory is probed for a non-trivial fixed point
\cite{Dorey:1990sz}, for
technicolor models \cite{Appelquist:1988ee},
gauged Nambu--Jona-Lasinio models \cite{Leung:1989hw},
and more recently  color superconductivity
\cite{Hong:2000fh}.  A similar quest  takes place in top-color models,
where the  mass of the top  quark is generated through  a gap equation
involving a strongly interacting massive gauge field
\cite{Bardeen:1990ds}.   The usual  conceptual drawback
is  that
sooner  or later  one is  forced  to choose  a gauge,  resorting to  a
variety of arguments; but gauge  choices cast in general doubts on the
robustness  of the conclusions  thusly reached. 

To address the problems of the gauge-dependence of
off-shell Green's functions a method known as the PT 
has been introduced \cite{Cornwall:1982zr,Cornwall:1989gv}.
The PT is  a
diagrammatic  method which exploits  the  underlying  symmetries
encoded  in  a  {\it  physical} amplitude  such  as  an  $S$-matrix  element,
or a Wilson loop,
in  order  to  construct effective   Green's   functions    with   special
properties.    The aforementioned symmetries,  even though they are  always
present, they are  usually concealed by  the gauge-fixing  procedure.
The  PT
makes them  manifest by means  of a  fixed algorithm,  which does  {\it not}
depend on  the gauge-fixing scheme  one uses in order  to quantize the theory,
{\it i.e.} regardless of the  set of Feynman rules used when writing down  the
$S$-matrix  element.   The method  exploits  the elementary  Ward
identities
triggered  by the  longitudinal  momenta appearing  inside  Feynman  diagrams
in  order  to  enforce massive cancellations. The realization of these
cancellations  mixes non-trivially contributions stemming from diagrams of
different  kinematic nature (propagators, vertices, boxes). Thus,      a  given
physical  amplitude  is reorganized into sub-amplitudes,   which  have   the
same   kinematic   properties  as conventional $n$-point functions and, in
addition, are  endowed with  desirable physical  properties, such as
GFP-independence.
Finally, the PT amounts to a non-trivial
{\it reorganization of the perturbative expansion}
The role of the PT  when dealing with SD equations is to
(eventually)  trade   the  conventional  SD   series  for
another,  written  in terms  of  the  new, gauge-independent  building
blocks \cite{Cornwall:1982zr,Mavromatos:1999jf,Sauli:2002tk}
The upshot  of this program would then
be  to truncate this  new series,  by keeping  only a  few terms  in a
``dressed-loop'' expansion, and maintain exact gauge-invariance, while
at the same time accommodating non-perturbative effects.
We hasten to emphasize that the aforementioned program has {\it not} 
been completed; however, a great deal of important insight 
on the precise 
GFP-cancellation mechanism has been accumulated, and the 
field-theoretic properties of gauge-independent Green's 
functions have been established in detail.

\medskip

\centerline{\bf \large{An explicit one-loop example}.}

\medskip

We next explain how the PT
gives  rise to
effective,  gauge-independent  fermion  self-energies at one-loop,
for the  case of  QED and  QCD \cite{Binosi:2001hy}. 
As will become clear in what follows, the procedure described 
does {\it not} depend on the dimensionality of space-time;
in particular, it applies unaltered at $d=3,4$. 
We will assume that the theory has been  gauge-fixed by introducing in the
gauge-invariant Lagrangian a gauge-fixing term of the form
$\frac{1}{2\xi}(\partial_{\mu}A^{\mu})^2$, {\it i.e.}
 a  linear, covariant gauge;
the parameter $\xi$ is the GFP. This gauge-fixing term gives rise to a bare
gauge-boson propagator of the form
\begin{equation}
\Delta_{\mu\nu}(\ell,\xi ) =
-\frac{\D i}{\D \ell^2}
\left[\ g_{\mu\nu} - (1-\xi) \frac{\D \ell_\mu
\ell_\nu}{\D \ell^2}\right]
\end{equation}
which explicitly depends on $\xi$.  The trivial  color factor $\delta_{ab}$
appearing in the (gluon) propagator  has been suppressed. The form of
$\Delta_{\mu\nu}(\ell,\xi )$ for the  special choice   $\xi =1$ (Feynman gauge)
will be of central importance in what follows; we will denote it by
$\Delta_{\mu\nu}^{F}(\ell)$, {\it i.e.}
\begin{equation}
\Delta_{\mu\nu}(\ell, 1 )\equiv \Delta_{\mu\nu}^{F}(\ell)
=  -\frac{\D i}{\D \ell^2}\, g_{\mu\nu}\,  .
\end{equation}
$\Delta_{\mu\nu}(\ell,\xi )$  and $\Delta_{\mu\nu}^{F}(\ell)$ will  be
denoted graphically   as  follows:

\begin{center}
\begin{picture}(0,20)(100,-15)

\Gluon(-5,-5)(30,-5){2.5}{7} \Photon(160,-5)(195,-5){2}{6}

\Text(35,-5)[l]{\normalsize{$\equiv   i  \Delta_{\mu\nu}(\ell,\xi)$,}}
\Text(200,-5)[l]{\normalsize{$\equiv i \Delta_{\mu\nu}^F(\ell)$.}}

\end{picture}
\end{center}
For the diagrammatic proofs that   will follow, in  addition  to the
propagators   $\Delta_{\mu\nu}(\ell)$  and   $\Delta_{\mu\nu}^F(\ell)$
introduced  above,   we   will need  six   auxiliary  propagator-like
structures, as shown below:

\begin{center}
\begin{picture}(0,80)(100,-75)

\Photon(-5,-5)(30,-5){2}{6}
\Photon(-5,-3.5)(30,-3.5){2}{6}
\Text(11,-5)[c]{\rotatebox{19}{\bf{\big /}}}
\Text(10.4,-5)[c]{\rotatebox{19}{\bf{\big /}}}
\Text(14,-5)[c]{\rotatebox{19}{\bf{\big /}}}
\Text(14.6,-5)[c]{\rotatebox{19}{\bf{\big /}}}
\Text(35,-5)[l]{\normalsize{$\equiv\ \frac{\D \ell_\mu
\ell_\nu}{\D \ell^4}$}}

\Photon(-5,-35)(30,-35){2}{6}
\Photon(-5,-33.5)(30,-33.5){2}{6}
\Text(12.2,-35)[c]{\rotatebox{19}{\bf{\big /}}}
\Text(12.8,-35)[c]{\rotatebox{19}{\bf{\big /}}}
\Vertex(-5,-35){2}
\Text(35,-35)[l]{\normalsize{$\equiv\ \frac{\D \ell_\mu}{\D \ell^4}$}}

\Photon(-5,-65)(30,-65){2}{6}
\Photon(-5,-63.5)(30,-63.5){2}{6}
\Vertex(-5,-65){2}
\Vertex(30,-65){2}
\Text(35,-65)[l]{\normalsize{$\equiv\ \frac{\D 1}{\D \ell^4}$}}

\Photon(160,-5)(195,-5){2}{6}
\Text(176,-5)[c]{\rotatebox{19}{\bf{\big /}}}
\Text(175.4,-5)[c]{\rotatebox{19}{\bf{\big /}}}
\Text(179,-5)[c]{\rotatebox{19}{\bf{\big /}}}
\Text(179.6,-5)[c]{\rotatebox{19}{\bf{\big /}}}
\Text(200,-5)[l]{$\equiv\ \frac{\D \ell_\mu
\ell_\nu}{\D \ell^2}$}

\Photon(160,-35)(195,-35){2}{6}
\Vertex(160,-35){2}
\Text(200,-35)[l]{\normalsize{$\equiv\ \frac{\D \ell_\mu}{\D \ell^2}$}}
\Text(177.2,-35)[c]{\rotatebox{19}{\bf{\big /}}}
\Text(177.8,-35)[c]{\rotatebox{19}{\bf{\big /}}}

\Photon(160,-65)(195,-65){2}{6}
\Vertex(160,-65){2}
\Vertex(195,-65){2}
\Text(200,-65)[l]{\normalsize{$\equiv\ \frac{\D 1}{\D \ell^2}$}}

\end{picture}
\end{center}
All of these six structures will arise from  algebraic manipulations of the
original $\Delta_{\mu\nu}(\ell)$. For example, in terms of the above notation
we have the following simple relation (we will set
$\lambda  \equiv \xi-1$):

\begin{center}
\begin{picture}(0,10)(60,20)

\Gluon(5,25)(40,25){2.5}{7}
\Text(45,25)[l]{\normalsize{$\equiv$}}
\Photon(60,25)(95,25){2}{6}
\Text(100,25)[l]{\normalsize{$+\ \lambda$}}
\Photon(120,25.75)(155,25.75){2}{6}
\Photon(120,24.25)(155,24.25){2}{6}
\Text(136,25)[c]{\rotatebox{19}{\bf{\big /}}}
\Text(135.4,25)[c]{\rotatebox{19}{\bf{\big /}}}
\Text(139,25)[c]{\rotatebox{19}{\bf{\big /}}}
\Text(139.6,25)[c]{\rotatebox{19}{\bf{\big /}}}

\end{picture}
\end{center}
We next turn to the study of the gauge-dependence of the fermion self-energy
(electron in QED, quarks in QCD). The inverse electron propagator  of order $n$
in the perturbative expansion has the form (again suppressing color)
\begin{equation}
S_n^{-1}(p,\xi) = \pslush -m - \Sigma^{(n)}(p,\xi)
\end{equation}
where $\Sigma^{(n)} (p,\xi)$ is the $n-th$ order  self-energy. Clearly $
\Sigma^{(0)} = 0$, and $S_0^{-1}(p) = \pslush -m$. The quantity $\Sigma^{(n)}
(p,\xi)$ depends  explicitly on $\xi$ already for $n=1$. In particular

\be
\Sigma^{(1)}(p,\xi) = \int  [d\ell]
\gamma^{\mu} S_0(p+\ell) \gamma^{\nu}
\Delta_{\mu\nu}(\ell,\xi) =
\Sigma^{(1)}_F (p)
+ \lambda \Sigma^{(1)}_L (p)
\label{STOT}
\ee
with
\begin{equation}
\Sigma^{(1)}_F (p) \equiv \Sigma^{(1)}(p,1) =
\int  [d\ell]
\gamma^{\mu} S_0 (p+\ell) \gamma^{\nu}
\Delta_{\mu\nu}^F (\ell)
\label{SF}
\end{equation}
and
\begin{eqnarray}
\Sigma^{(1)}_L (p)
&=& - S_0^{-1}(p)
\int  \frac{[d\ell]}{\ell^4} \, S_0(p+\ell)\gamma^{\nu} \ell_{\nu}
= - \int \frac{[d\ell]}{\ell^4} \,   \ell_{\mu}\gamma^{\mu}
S_0(p+\ell)\,\,  S_0^{-1}(p)
\nonumber\\
&=&
  S_0^{-1}(p) \, \int \frac{[d\ell]}{\ell^4} S_0(p+\ell) \,\, S_0^{-1}(p)
 - S_0^{-1}(p) \int  \frac{[d\ell]}{\ell^4} .
\label{SL}
\end{eqnarray}
In the above formulas  $ [d\ell] \equiv g^2
\mu^{2\epsilon}\frac{d^D \ell}{(2\pi)^D}$,
with $D=4-2\epsilon$ the dimension of space-time, $\mu$ the 't Hooft
mass, and $g$ the gauge coupling ($ g\equiv e$ for
QED, and $ g\equiv g_s$ for QCD). The subscripts ``F'' and ``L'' stand for
``Feynman'' and  ``Longitudinal'', respectively.  Notice that $\Sigma^{(1)}_L$
is proportional to  $S_0^{-1}(p)$ and thus vanishes ``on-shell''.  The most
direct way to arrive at the results of Eq.(\ref{SL}) is to employ the
fundamental WI
\begin{equation}
\elle=S_0^{-1}(p+\ell)-S_0^{-1}(p),
\label{WIzero}
\end{equation}
which is triggered every time the longitudinal
momenta of $\Delta_{\mu\nu}(\ell,\xi)$ gets contracted with the appropriate
$\gamma$ matrix appearing in the  vertices.
Diagrammatically, this elementary WI gets translated to

\begin{center}
\begin{picture}(0,40)(60,20)

\Line(0,25)(50,25)
\Photon(24.3,50)(24.3,25){2}{5}
\Photon(25.7,50)(25.7,25){2}{5}
\Text(25,37.2)[c]{\rotatebox{-71}{\bf{\big /}}}
\Text(25,37.8)[c]{\rotatebox{-71}{\bf{\big /}}}
\Text(55,25)[l]{$\equiv$}

\Line(70,25)(100,25)
\Photon(100.7,25)(100.7,50){2}{5}
\Photon(99.3,25)(99.3,50){2}{5}
\Vertex(100,25){2}

\Text(120,25)[c]{$-$}

\Line(140,25)(170,25)
\Photon(140.7,25)(140.7,50){-2}{5}
\Photon(139.3,25)(139.3,50){-2}{5}
\Vertex(140,25){2}

\end{picture}
\end{center}
Then, the diagrammatic representation of
Eq.(\ref{STOT}), Eq.(\ref{SF}), and Eq.(\ref{SL}) will be given by

\begin{equation}
\begin{picture}(0,110)(130,-45)

\Line(-15,25)(55,25)
\GlueArc(20,25)(20,0,180){2.5}{9}
\Text(60,25)[l]{$\equiv$}

\Line(70,25)(140,25)
\PhotonArc(105,25)(20,0,180){2}{7.5}
\Text(145,25)[l]{$-\ \lambda$}

\Line(170,25)(230,25)
\PhotonArc(190,25)(20,0,180){2}{7.5}
\PhotonArc(190,25)(18.8,0,180){2}{7.5}
\Text(190.9,44.5)[c]{\rotatebox{19}{\bf{\big /}}}
\Text(190.3,44.5)[c]{\rotatebox{19}{\bf{\big /}}}

\Text(60,-30)[l]{$=$}
\Line(70,-30)(140,-30)
\PhotonArc(105,-30)(20,0,180){2}{7.5}
\Text(145,-30)[l]{$+\ \lambda$}

\Line(170,-30)(230,-30)
\PhotonArc(200,-40)(30,20,160){2}{7.5}
\PhotonArc(200,-40)(28.8,20,160){2}{7.5}
\Vertex(170,-30){2}
\Vertex(230,-30){2}
\Text(237,-30)[l]{$-\ \lambda$}

\Line(270,-30)(320,-30)
\PhotonArc(270,-20)(10,180,540){2}{9}
\PhotonArc(270,-20)(8.8,180,540){2}{9}

\Vertex(270,-30){2}
\Vertex(170,25){2}

\label{GR}
\end{picture}
\end{equation}

When considering physical amplitudes,  the characteristic structure of the
longitudinal parts established above   allows for their cancellation against
identical contributions  originating from diagrams which are kinematically
different from fermion self-energies,  such as vertex-graphs or boxes, {\it
without} the need for integration over the internal virtual momenta.  This last
property is important because  in this way the original  kinematical identity
is guaranteed to be maintained; instead, loop integrations generally mix the
various kinematics.   Diagrammatically, the action of the WI  is very
distinct:  it always gives rise to unphysical effective vertices, {\it i.e.}
vertices
which do not appear in the original Lagrangian; all such vertices   cancel in
the full, gauge-invariant amplitude.

\begin{figure}[t]

\begin{picture}(100,100)(-60,0)

\Text(20,25)[l]{\footnotesize{(a)}}
\Text(135,25)[l]{\footnotesize{(b)}}
\Text(250,25)[l]{\footnotesize{(c)}}

\Text(24,70)[l]{\scriptsize{$k$}}
\Text(71,70)[l]{\scriptsize{$k+\ell$}}
\Text(71,30)[l]{\scriptsize{$k+\ell-q$}}

\Text(165,81)[c]{\scriptsize{$k$}}
\Text(192,40)[l]{\scriptsize{$k-q$}}
\Text(165,19)[c]{\scriptsize{$k+\ell-q$}}

\Text(280,81)[c]{\scriptsize{$k+\ell$}}
\Text(250,60)[l]{\scriptsize{$k$}}
\Text(280,19)[c]{\scriptsize{$k-q$}}

\CArc(20,50)(5,0,360)
\Line(18,52)(22,48)
\Line(22,52)(18,48)
\CArc(50,50)(25,0,360)
\CArc(80,50)(5,0,360)
\Line(78,52)(82,48)
\Line(82,52)(78,48)

\CArc(135,50)(5,0,360)
\Line(133,52)(137,48)
\Line(137,52)(133,48)
\CArc(165,50)(25,0,360)
\CArc(195,50)(5,0,360)
\Line(193,52)(197,48)
\Line(197,52)(193,48)

\CArc(250,50)(5,0,360)
\Line(248,52)(252,48)
\Line(252,52)(248,48)
\CArc(280,50)(25,0,360)
\CArc(310,50)(5,0,360)
\Line(308,52)(312,48)
\Line(312,52)(308,48)

\Gluon(50,25)(50,75){2}{9}

\GlueArc(165,25)(19.2,22.5,157.5){2}{9}

\GlueArc(280,75)(19.2,202.5,337.5){2}{9}

\end{picture}

\caption{One loop diagram contributing to the QED/QCD fermion self-energy.}

\label{fig1}
\end{figure}
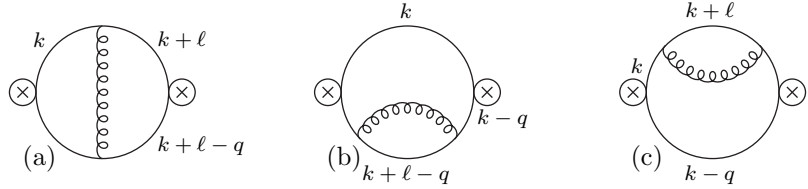

To actually pursue these special
cancellations explicitly one may choose among a variety
of gauge invariant quantities.  For example, one may consider the current
correlation function  $I_{\mu\nu}$ defined as (in momentum space)
\be
I_{\mu\nu}(q) = i \int d^{4}x e^{iq\cdot x}
\langle 0  | T \left[J_{\mu}(x) J_{\nu}(0) \right] |0 \rangle
 = (g_{\mu\nu} q^2 - q_{\mu}q_{\nu}) I(q^2) \, ,
\label{IQED}
\ee
where the current $J_{\mu} (x)$ is given by $J_{\mu}(x) =\ :\!\bar{Q} (x)
\gamma_{\mu} Q (x)\!:$. $I_{\mu\nu}(q)$  coincides with the photon
vacuum polarization of QED.

To see explicitly the mechanism enforcing  these cancellations in the   QED and
QCD cases, we first  consider the one-loop photonic or gluonic corrections,
respectively, to the quantity  $I_{\mu\nu}$. Clearly either set of corrections
is GFP-independent, since the current $J_{\mu}(x)$ is invariant under both the
$U(1)$ and the $SU(3)$ gauge transformations.

The relevant diagrams are those shown in Fig.\ref{fig1}. To see the
appearance of the unphysical vertices,
we carry out the manipulations presented
in Eq.(\ref{STOT}), Eq.(\ref{SF}), and Eq.(\ref{SL}), or, equivalently, in
 Eq.(\ref{GR}),
this time  embedded inside $I_{\mu\nu}(q)$. Thus, from  diagrams (b) and
(c) we arrive at
\begin{center}
\begin{picture}(100,50)(30,25)

\Text(-55,50)[l]{(b)+(c) $\to\ 2\lambda$}

\CArc(20,50)(5,0,360)
\Line(18,52)(22,48)
\Line(22,52)(18,48)
\CArc(50,50)(25,0,360)
\CArc(80,50)(5,0,360)
\Line(78,52)(82,48)
\Line(82,52)(78,48)
\PhotonArc(50,25)(19.2,22.5,157.5){2}{9}
\PhotonArc(50,25)(18,22.5,157.5){2}{9}
\Text(48.5,44.2)[c]{\rotatebox{19}{\bf{\big /}}}
\Text(47.9,44.2)[c]{\rotatebox{19}{\bf{\big /}}}
\Text(51.5,44.2)[c]{\rotatebox{19}{\bf{\big /}}}
\Text(52.1,44.2)[c]{\rotatebox{19}{\bf{\big /}}}

\Text(90,50)[l]{$=-\,2\lambda$}
\hspace{-1.3 cm}

\CArc(160,50)(5,0,360)
\Line(158,52)(162,48)
\Line(162,52)(158,48)
\CArc(190,50)(25,0,360)
\CArc(220,50)(5,0,360)
\Line(218,52)(222,48)
\Line(222,52)(218,48)
\PhotonArc(215,25)(25.7,90,180){-2}{7.5}
\PhotonArc(215,25)(24.3,90,180){-2}{7.5}
\Vertex(215,50){2}
\Text(196,41)[c]{\rotatebox{65}{\bf{\big /}}}
\Text(196.45,41.45)[c]{\rotatebox{65}{\bf{\big /}}}

\end{picture}
\end{center}
We thus see that since the action of the
elementary WI of  Eq.(\ref{WIzero}) amounts to the cancellation of internal
propagators,
its diagrammatic
consequence
is that of introducing an unphysical effective vertex,
describing an interaction of the form $\gamma\gamma \bar{Q} Q$ or
$\gamma G \bar{Q} Q$, depending on whether we consider photonic or
gluonic corrections.
This type of vertex
may be depicted by means of a Feynman rule of the form

\begin{center}
\begin{picture}(0,35)(20,10)

\CArc(20,10)(5,0,360)
\Line(18,12)(22,8)
\Line(22,12)(18,8)

\Line(-5,15)(45,15)
\Gluon(20,15)(20,40){2}{5}
\Vertex(20,15){2}

\Text(50,15)[l]{$\equiv i\gamma_\mu$}
\Text(30,10)[c]{\footnotesize{$\mu$}}

\end{picture}
\end{center}
being $\mu$ the index of the external current.

To see how the above unphysical contributions cancel inside  $I_{\mu\nu}$
we turn to diagram (a). The action of
the WI may be
translated to the following diagrammatic picture

\begin{center}
\begin{picture}(100,50)(85,25)

\Text(-30,50)[l]{(a) $\to\ \lambda$}

\CArc(20,50)(5,0,360)
\Line(18,52)(22,48)
\Line(22,52)(18,48)
\CArc(50,50)(25,0,360)
\CArc(80,50)(5,0,360)
\Line(78,52)(82,48)
\Line(82,52)(78,48)
\Photon(49.25,25)(49.25,75){2}{9}
\Photon(50.75,25)(50.75,75){2}{9}
\Text(50,51.5)[c]{\rotatebox{-71}{\bf{\big /}}}
\Text(50,52.1)[c]{\rotatebox{-71}{\bf{\big /}}}
\Text(50,48.5)[c]{\rotatebox{-71}{\bf{\big /}}}
\Text(50,47.9)[c]{\rotatebox{-71}{\bf{\big /}}}
\Text(90,50)[l]{$=\ \ \lambda$}
\hspace{-1.5 cm}

\Text(155,25)[lb]{\footnotesize{$(\alpha)$}}

\CArc(160,50)(5,0,360)
\Line(158,52)(162,48)
\Line(162,52)(158,48)
\CArc(190,50)(25,0,360)
\CArc(220,50)(5,0,360)
\Line(218,52)(222,48)
\Line(222,52)(218,48)
\PhotonArc(215,25)(25.7,90,180){-2}{7.5}
\PhotonArc(215,25)(24.3,90,180){-2}{7.5}
\Vertex(215,50){2}
\Text(196,41)[c]{\rotatebox{65}{\bf{\big /}}}
\Text(196.45,41.45)[c]{\rotatebox{65}{\bf{\big /}}}
\Text(230,50)[l]{$-\ \ \lambda$}
\hspace{-1.6 cm}

\Text(295,25)[lb]{\footnotesize{$(\beta)$}}

\CArc(300,50)(5,0,360)
\Line(298,52)(302,48)
\Line(302,52)(298,48)
\CArc(330,50)(25,0,360)
\CArc(360,50)(5,0,360)
\Line(358,52)(362,48)
\Line(362,52)(358,48)
\PhotonArc(305,25)(25.7,0,90){2}{7.5}
\PhotonArc(305,25)(24.3,0,90){2}{7.5}
\Vertex(305,50){2}
\Text(324,41)[c]{\rotatebox{-25}{\bf{\big /}}}
\Text(323.6,41.4)[c]{\rotatebox{-25}{\bf{\big /}}}

\end{picture}
\end{center}
It is then elementary to establish that the two diagrams on the
right-hand side
of the
above diagrammatic  equation add up.

Summing up the two equations above, it is clear how  the gauge dependent part
of the one loop amplitude cancel completely.
Having proved that the GFP-dependent contributions coming from the
original graphs containing $\Sigma^{(1)}(p,\xi)$, {\it i.e.}
Fig.\ref{fig1}(b) and Fig.\ref{fig1}(c) cancel exactly against
equal but opposite {\it propagator-like} contributions coming from
Fig.\ref{fig1}(a), one is left with the ``pure'' GFP-independent
one-loop fermion self-energy, $\widehat{\Sigma}^{(1)}(p)$. Clearly,
it coincides with the $\Sigma^{(1)}_F (p)$ of Eq.(\ref{SF}), i.e.
\begin{equation}
\widehat{\Sigma}^{(1)}(p) = \Sigma^{(1)}_F (p).
\label{GFPI}
\end{equation}

\medskip

\centerline{\bf \large{All-order results}.}

\medskip
 
\begin{figure}[!t]
\begin{center}
\includegraphics[width=10cm]{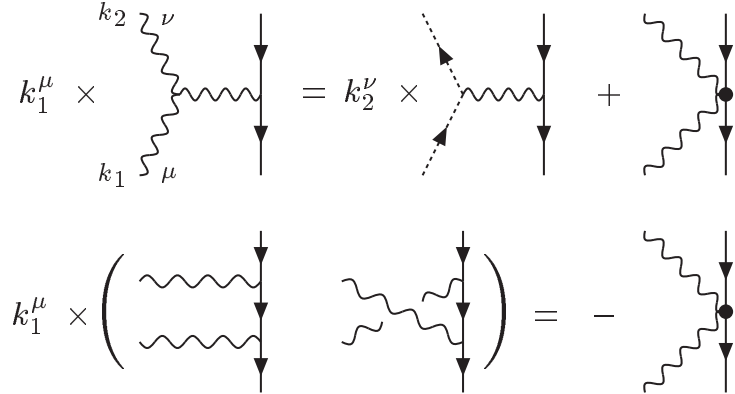}
\end{center} 
\caption{\label{fig:1} The tree-level version
of the fundamental $s$-$t$ channel cancellation.}
\end{figure}

The generalization of the PT  to all orders has been
recently accomplished \cite{Binosi:2002ft,Binosi:2003rr}. 
The main points are the following:

{\bf{(i)}} In a fully non-Abelian context the longitudinal momenta
responsible for the various rearrangements between different 
Green's functions do not stem solely from the tree-level 
propagator of the gauge boson (as in the previous one-loop example) 
but also from the vertices carrying momenta, i.e. the three-boson vertex 
$\Gamma^{eab,[0]}_{\alpha\mu\nu}$ \cite{Cornwall:1989gv}.  
It turns out that this latter type of longitudinal momenta  
triggers a fundamental
cancellation taking place between  graphs of distinct kinematic nature
($s$-channel versus $t$-channel graphs), shown 
to lowest order in \ref{fig:1}. In particular, 
when  the  $s$-channel and  $t$-channel 
diagrams  of Fig.\ref{fig:1})  
are contracted by
a  common  longitudinal momentum,  one  obtains  from  either graph  a
common,  propagator-like part,  which eventually  cancels  against the
other \cite{Papavassiliou:1996zn}.
These parts  display  the characteristic  feature that,  when
depicted  by  means  of  Feynman  diagrams,  they  contain  unphysical
vertices (Fig.\ref{fig:1}), {\it i.e.}, vertices which do not exist in
the original Lagrangian, and cancel in any observable quantity.

{\bf{(ii)}} The generalization of the above cancellation to all orders 
in perturbation theory has been demonstrated recently in 
\cite{Binosi:2002ft,Binosi:2003rr}.
In the all-order case one considers 
the divergence  
of the four-point function $A_{\mu}^a\,
A_{\nu}^b\, q^i\, \bar{q}^j$, with the gluons $A^a_{\mu}$, $A^b_{\nu}$
off-shell,   and  the  quarks   $q^i$,  $\bar{q}^j$   on-shell. 
The aforementioned
four-point function constitutes a common kernel to all self-energy and
vertex  diagrams  appearing in  the  process  $q^m  \bar{q}^n \to  q^i
\bar{q}^j$
(note that the diagrams of Fig.\ref{fig:1} are simply the tree-level  
contribution  to the amplitude $A_{\mu}^a\, A_{\nu}^b\, q^i\, \bar{q}^j$). 
 As  has  been  shown  in
\cite{Binosi:2002ft,Binosi:2003rr} the  judicious exploitation of the
all-order Slavnov-Taylor identity
that this
Green's function satisfies allows  for the all-order generalization of
the PT procedure.

{\bf{(iii)}}
The effective gauge-independent Green's function constructed by the
PT coincide to {\it all orders} with the the background field method
Green's functions when the latter are computed in the
(quantum) Feynman gauge \cite{Binosi:2002ft,Binosi:2003rr}.

{\bf{(iv)}} The equality of
Eq~(\ref{GFPI}) persists to {\it all orders}: 
after all gauge-cancellation have been carried out, 
the gauge-independent PT fermion self-energy coincides
with the conventional one computed in the renormalizable Feynman gauge. 
The general statement of {\bf{(iii)}} is of course valid in this case as well,
simply because  the fermion self-energy computed in the renormalizable Feynman 
gauge happens to be identical (to all orders) to  
the fermion self-energy computed in the 
background field method Feynman gauge. 
This last statement is of course not true in general; Green's 
functions computed in the  renormalizable Feynman gauge do not 
coincide with the corresponding  Green's functions computed 
in the background field method Feynman gauge. What is always true 
however is {\bf{(iii)}}.

{\bf{(v)}} 
In QED (but not in QCD) 
the statement of {\bf{(iv)}} is true also for the 
off-shell photon-fermion vertex $\Gamma_{\mu}$.

\end{document}